\documentclass[final,3p,times,authoryear]{elsarticle}
\usepackage{amssymb}
\usepackage{amsmath}
\usepackage{xcolor}
\DeclareMathOperator{\sgn}{sgn}
\usepackage{hyperref}
\journal{Journal of Theoretical Biology}

\begin{document}

\begin{frontmatter}



\title{A coupled neural field model for the standard consolidation theory}


\author[inst1,inst2]{Lisa Blum Moyse\fnref{label2}\corref{cor1}}
\cortext[cor1]{Corresponding author}
\fntext[label2]{Present adress: University of Konstanz, Cluster for the Advanced Study of Collective Behaviour, Konstanz, 78457, Germany}

\affiliation[inst1]{organization={LIRIS, CNRS UMR 5205},%Department and Organization
            % addressline={Address One}, 
            city={Villeurbanne},
            postcode={F-69621}, 
            country={France}}
\ead{lisa.blum-moyse@uni-konstanz.de}
\author[inst2]{Hugues Berry}
% \author[inst1,inst2]{Author Three}

\affiliation[inst2]{organization={AIstroSight, Inria, Hospices Civils de Lyon, Universite Claude Bernard Lyon 1},%Department and Organization
            city={Villeurbanne},
            postcode={F-69603}, 
            country={France}}
\ead{hugues.berry@inria.fr}

\begin{abstract}
The standard consolidation theory states that short-term memories located in the hippocampus enable the consolidation of long-term memories in the neocortex. In other words, the neocortex slowly learns long-term memories with a transient support of the hippocampus that quickly learns unstable memories. However, it is not clear yet what could be the neurobiological mechanisms underlying these differences in learning rates and memory time-scales. Here, we propose a novel modelling approach of the standard consolidation theory, that focuses on its potential neurobiological mechanisms. In addition to synaptic plasticity and spike frequency adaptation, our model incorporates adult neurogenesis in the dentate gyrus as well as the difference in size between the neocortex and the hippocampus, that we associate with distance-dependent synaptic plasticity. We also take into account the interconnected spatial structure of the involved brain areas, by incorporating the above neurobiological mechanisms in a coupled neural field framework, where each area is represented by a separate neural field with intra- and inter-area connections. To our knowledge, this is the first attempt to apply neural fields to this process. Using numerical simulations and mathematical analysis, we explore the short-term and long-term dynamics of the model upon alternance of phases of hippocampal replay and retrieval cue of an external input. This external input is encodable as a memory pattern in the form of a multiple bump attractor pattern in the individual neural fields. In the model, hippocampal memory patterns become encoded first, before neocortical ones, because of the smaller distances between the bumps of the hippocampal memory patterns. As a result, retrieval of the input pattern in the neocortex at short time-scales necessitates the additional input delivered by the memory pattern of the hippocampus. Neocortical memory patterns progressively consolidate at longer times, up to a point where their retrieval does not need the support of the hippocampus anymore. At longer times, perturbation of the hippocampal neural fields by neurogenesis erases the hippocampus pattern, leading to a final state where the memory pattern is exclusively evoked in the neocortex. Therefore, the dynamics of our model successfully reproduces the main features of the standard consolidation theory. This suggests that neurogenesis in the hippocampus and distance-dependent synaptic plasticity coupled to synaptic depression and spike frequency adaptation, are indeed critical neurobiological processes in memory consolidation.

\end{abstract}


% \begin{highlights}
% \end{highlights}

\begin{keyword}
multi-bump attractors \sep long-term plasticity \sep systems consolidation memory \sep spike frequency adaptation \sep synaptic depression
% %% PACS codes here, in the form: \PACS code \sep code
% %% MSC codes here, in the form: \MSC code \sep code
% %% or \MSC[2008] code \sep code (2000 is the default)
\end{keyword}

\end{frontmatter}


\section{Introduction}
\label{introduction}
%\subsection{Standard consolidation theory}

Memories are believed to be biologically encoded as physical traces in the brain, or engrams. It is assumed that these engrams form through the strengthening of synaptic connections in neuronal ensembles, i.e. populations of neurons involved in a memory representation or a computational task~\citep{josselynFindingEngram2015}.
The process through which recently encoded memories are converted into more stable long-term stored information is referred to as consolidation~\citep{klinzingMechanismsSystemsMemory2019}. This term includes two mechanisms that occur at distinct spatial and time scales: synaptic and systems consolidation.
Synaptic consolidation is achieved through fast mechanisms (a few hours), like long-term synaptic plasticity~\citep{martinSynapticPlasticityMemory2000}.
The most common form of long-term synaptic plasticity is qualified as Hebbian, it strengthens synapses
between neurons that fire simultaneously, which stabilizes a memory trace. These processes are themselves embodied into wider and more systemic change processes invoked by systems consolidation. This second type of consolidation describes the gradual reorganization of memory patterns across different brain areas, which can endure weeks, months, or even years~\citep{klinzingMechanismsSystemsMemory2019}.

The theories of systems consolidation were introduced following neuropsychological observations of memory impairments in patients with medial temporal lobe (MTL) damages. The MTL includes the hippocampus and adjacent neocortical areas (perirhinal, entorhinal, parahippocampal)~\citep{squireMedialTemporalLobe2004}.
MTL lesions induce anterograde amnesia that is, the inability to form new memories. This disability concerns in most cases declarative memories, memories that can be consciously recalled~\citep{squireRetrogradeAmnesiaMemory1995,squireMedialTemporalLobe2004}. These studies introduced the idea that there exist different memory systems associated with different brain areas.
MTL injuries are also responsible for temporally-graded retrograde amnesia for declarative memories. Temporally-graded retrograde amnesia refers to a forgetting of memories encoded in the past, where the loss is more important for recent events.
In general, the wider the damages, the more ancient the erased memories: amnesia usually affects memories ranging from 1 to 2 years in the past when lesions are restricted to the hippocampus, to a situation where all past memories are forgotten when the damages include the whole MTL and the surrounding cortical regions. 
These studies indicate that recent declarative memories depend initially on the MTL, and are subsequently transferred to a durable storage place, possibly cortical areas.
Since then, many animal studies have examined the impact of hippocampal and cortical lesions on memory~\citep{franklandOrganizationRecentRemote2005}.

\begin{figure}
\centering
\includegraphics[width=0.95\textwidth]{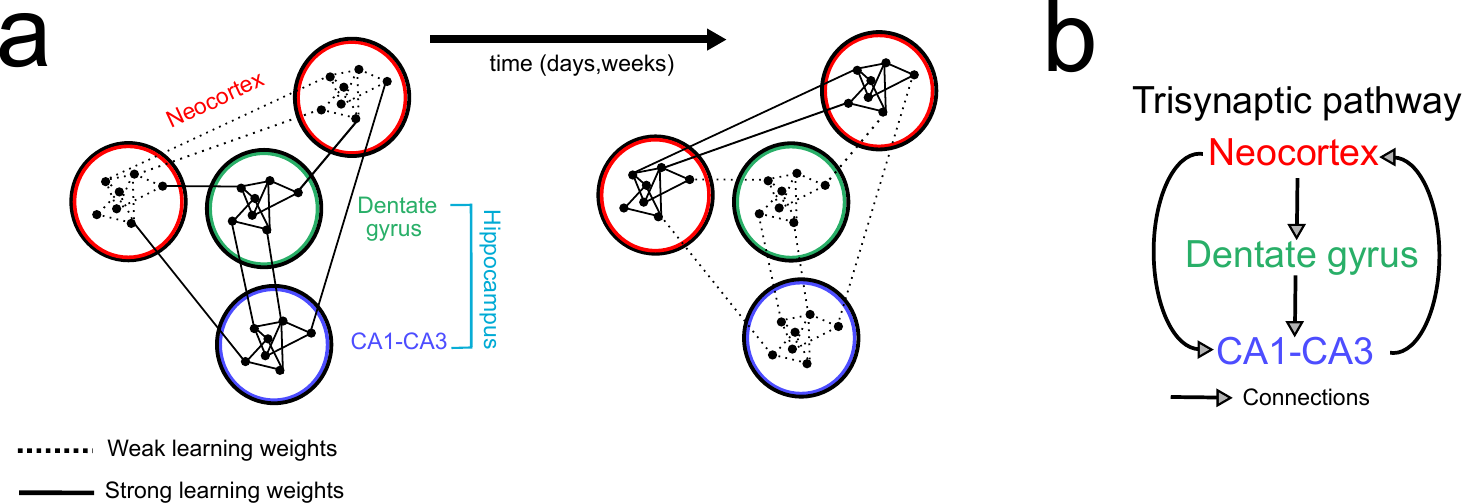}
\caption[Interactions between the different brain areas involved in the standard consolidation theory]{Interactions between the different brain areas involved in the standard consolidation theory. \textbf{(a)} Standard consolidation theory.  Initially, the engram is initially present in neocortical areas (red), in a weak form (i.e., not stabilized), and in the hippocampus (dentate gyrus in green and CA1-CA3 subfields in blue), in a stable form.
After some days the distribution of the memory is reorganized. Connections are consolidated in the neocortex, while neocortico-hippocampal and hippocampal connections are degraded. 
\textbf{(b)} A simplified scheme of the trisynaptic pathway, the circuit considered in the present model. The arrows represent the main synaptic connections between the three brain areas: neocortex, dentate gyrus, and hippocampus sub-fields CA1-CA3.}
\label{intro}
\end{figure}

In 1971, the initial studies on MTL damages led Marr to propose the first computational model describing systems consolidation~\citep{marrSimpleMemoryTheory1991}. In this model, it is suggested that a new event is quickly encoded in the hippocampus and that with time this memory will be progressively ``transferred'' to the neocortex, through repetitive replays of the patterns during sleep. Following this pioneering work, a series of connectionist models have conceptualized the features of the standard consolidation theory and produced results in coherence with neuropsychological observations~\citep{squireRetrogradeAmnesiaMemory1995,mcclellandWhyThereAre1995,kaliOfflineReplayMaintains2004,meeterTracelinkModelConsolidation2005,amaralSynapticReinforcementbasedModel2008,helferComputationalModelSystems2020a,howardModelBidirectionalInteractions2022}. These ideas have led to a contemporary model of systems memory consolidation, often referred to the standard consolidation theory (SCT)~\citep{weingartnerMemoryConsolidation2014,squireRetrogradeAmnesiaMemory1995} or the standard model of systems consolidation\footnote{We focus here on the widespread standard consolidation theory, although interesting alternative views exist, such as the multiple trace theory or the trace transformation theory, which will be addressed in the discussion.}, see figure~\ref{intro}~(a). According to this theory, a new experience and the set of features it contains are first encoded in different associative cortical modules. This information is then quickly transmitted and integrated by the hippocampus, which forms a compressed memory trace or pattern. Repeated offline reactivation of this hippocampal memory pattern can then activate the corresponding neocortical representations. Indeed the reactivation of memory patterns can occur during ``offline'' states, typically during sleep, or during ``online'' states, upon the encountering of an experience, or retrieval cue, related to this existing memory.
These replays result in a gradual strengthening of neocortical connections, which ultimately leads to a neocortical memory pattern incorporated with pre-existing local representations and solid enough to be independent of the hippocampus. Thus, the memory patterns are expected to have a short lifetime in the hippocampus, while the ones in the neocortex can remain encoded for years.
Therefore, on the one hand, the hippocampus is predicted to learn quickly online, and be essential for offline slow learning in the neocortex. According to previous connectionist models, this progressive neocortical learning would allow to prevent catastrophic interference and replacement of existing neocortical patterns with new ones~\citep{mcclellandWhyThereAre1995,roxinEfficientPartitioningMemory2013}. On the other hand, while the encoded memories are kept on the long term in the neocortex, they  are rapidly erased in the hippocampus. This clearance of hippocampal memories is believed to be important because of the limited storage capacity of the hippocampus~\citep{willshawMemoryModellingMarr2015}.

In spite of the abundant literature on systems consolidation and its general principles, it is still unclear what are the main neurobiological mechanisms involved and how these processes intervene in the differences in learning speed and memory stability between the hippocampus and the neocortex. In particular, to our knowledge, this issue has not been addressed in the previous mathematical models of SCT. Some of the published modelling studies directly assumed larger learning rates for hippocampal or cortico-hippocampal connections compared to neocortical ones~\citep{squireRetrogradeAmnesiaMemory1995,mcclellandWhyThereAre1995,kaliOfflineReplayMaintains2004,
meeterTracelinkModelConsolidation2005,amaralSynapticReinforcementbasedModel2008,helferComputationalModelSystems2020a,howardModelBidirectionalInteractions2022}. However, it seems unlikely that variations in the rates of long-term synaptic plasticity would induce differences in learning rates that are of the order of several days or weeks~\citep{lismanWhyCortexSlow2001a}. We believe that structural plasticity i.e., the plasticity of the synaptic wiring diagram could be a better candidate to explain learning rate differences~\citep{chklovskiiCorticalRewiringInformation2004}. Indeed, the neocortex is a large structure comprising a very large number of neurons with sparse interconnectivity: a neuron of the neocortex is physically connected to only a small fraction of its local neighbors~\citep{rollsNeuralNetworksBrain1997}. In contrast, the size of the hippocampus is much lower, which implies a smaller degree of sparsity and a larger fraction of pre-existing connections between hippocampal neurons. External stimulations can strengthen the pre-existing hippocampal synapses, leading to fast encoding. Such a fast encoding would be unlikely to happen in the much larger and sparser neocortical networks. Here, a new synaptic connection formed between two neocortical neurons would be stabilized with time through hippocampal replay via structural plasticity. This incremental process would result in memories that slowly consolidate in associative cortical modules~\citep{franklandACaMKIIdependentPlasticityCortex2001b,lismanWhyCortexSlow2001a}.

Beyond their reinforcement, little is understood about the clearance of memories~\citep{davisBiologyForgettingPerspective2017}. 
While regular hippocampal memories erasure seems a mandatory aspect in SCT models~\citep{mcclellandWhyThereAre1995,franklandOrganizationRecentRemote2005}, usually implemented as larger forgetting rates for hippocampal or cortico-hippocampal connections than for neocortical ones~\citep{squireRetrogradeAmnesiaMemory1995,mcclellandWhyThereAre1995,kaliOfflineReplayMaintains2004,
meeterTracelinkModelConsolidation2005,amaralSynapticReinforcementbasedModel2008,helferComputationalModelSystems2020a,howardModelBidirectionalInteractions2022}, its neurobiological origin has scarcely been investigated.
An interesting neurobiological hypothesis is that the dentate gyrus (DG) of the hippocampus, where neurogenesis occurs throughout life, would be involved in the gradual erasure of hippocampal memories~\citep{franklandHippocampalNeurogenesisForgetting2013,koNeurogenesisdependentTransformationHippocampal2021}. This effect can be explained by the perturbation of the hippocampal network stability as newborn neurons progressively integrate (over several weeks). 
Erasure is not likely to happen through the direct replacement of the pre-existing neurons of the engram by the newborn ones, since it seems that the neurons involved in an encoded pattern benefit from a survival advantage, so that they are likely to persist for longer times than the hippocampal engram~\citep{leunerLearningEnhancesSurvival2004}. Instead, erasure is thought to be due to local perturbation of the excitability of the pre-existing neurons by the newborn ones. Indeed, newborn cells are known to be highly excitable~\citep{geCriticalPeriodEnhanced2007,mongiatReliableActivationImmature2009}, so that their integration triggers offsetting mechanisms to restore local network homeostasis~\citep{karmarkarDifferentFormsHomeostatic2006}. Under these processes, the excitability of pre-existing neurons is lowered or their synapses weakened by synaptic scaling, which could progressively hinder the reactivation of hippocampal engrams.
This process has been implemented in mathematical models of hippocampal networks, confirming that the addition of new neurons to the DG layer induces memory degradation~\citep{deisserothExcitationNeurogenesisCouplingAdult2004,meltzerRoleCircuitHomeostasis2005,weiszNeurogenesisInterferesRetrieval2012}. 
%It seems that the fate of a memory is determined by the balance between the consolidation and decay processes: if a memory is never reactivated, it gradually disappears. Contrary to those in the hippocampus, neocortical memories, once consolidated, could perdure through online replays via retrieval cues. Indeed it is important to note that a retrieval cue will stimulate only a part of the neocortical pattern. That is why the neocortex is dependent on the hippocampus to retrieve its whole pattern at the beginning, while when all connections are established this neocortical engram can be retrieved independently (see figure~\ref{mainsteps}).\\

We present below a novel mathematical model of SCT that incorporates a set of potential neurobiological mechanisms. In addition to spike-frequency adaptation and synaptic plasticity, we assumed that the difference of connectivity structure of the neocortex and the hippocampus is playing a role in SCT, in particular the larger size and more sparse structure of the neocortex. We also incorporate neurogenesis in the hippocampus, assuming that this newborn neurons could play a role in memory erasure. Using numerical simulations and mathematical analyses, we explore the short and long term dynamics of the model along the iterated phases of a typical consolidation process.

%\subsection{Neural fields}
\section{Overview of the model}
Contrary to previous SCT connectionist models, we chose to represent the involved process with neural fields. To our knowledge, this is the first time that this framework is used to model SCT. This modelling approach assumes a continuum representation of large-scale biological neural networks. These models are composed of nonlinear integro-differential equations, where the integration kernels allow for spatial distributions of neural connections~\citep{bressloffSpatiotemporalDynamicsContinuum2012,coombesTutorialNeuralField2014c}. We believe that this framework is particularly suitable to model SCT for two reasons. 
First, the whole process involves a complex combination of mechanisms occurring at various time scales, that are challenging to capture with biophysically-detailed models, such as the Hodgkin-Huxley equations, that already describe several processes for a single neuron. Coarse-graining methods aim at producing simplified descriptions, derived from microscopic models. Thus, reduced models such as neural fields propose a good trade-off since they can express part of this complexity, while preserving some degree of mathematical analysis, even if these simplified models can become very abstract and depart from the initial microscopic description. One has to be careful with the conclusions raised using these models, but they can provide interesting indications.
Second, our approach consists in articulating SCT, which involves interactions at a tissue level, with the underlying neurobiological mechanisms. Neural fields, while modelling large areas, aim at staying close to biological observations, and have already proven their usefulness for the exploration of neurobiological processes such as epilepsy, encoding of visual stimuli, the representation of head direction or working memory~\citep{bressloffSpatiotemporalDynamicsContinuum2012}. Working memory corresponds to the temporary storage, on the time scale of seconds, of information in the brain. 
According to experimental reports, during a recall task, sensory inputs can be followed by a persistent activity within spatially defined clusters of neurons in the neocortex~\citep{goldman-rakicCellularBasisWorking1995}. 
This stationary pulse known as a bump of activity corresponds to a class of solutions in neural field theory and has also been studied for multiple bump patterns~\citep{laingMultipleBumpsNeuronal2002,ferreiraMultibumpSolutionsNeural2016}. 
Here, we propose that these bumps represent the different elements/sub-parts of an encoded memory pattern in SCT. However, in opposition to working memory, we consider here both short-term and long-term memories. This implies the introduction of synaptic plasticity in the neural field framework.\\

In its simplest form, the neural field equation is the following integro-differential equation~\citep{amariDynamicsPatternFormation1977}: 

\begin{equation}
\frac{\partial u}{\partial t}(x,t)=-u(x,t)+\int_{-\infty}^\infty dy w(x-y) f(u(y,t)-\kappa)+I_\text{ext}(x,t) 
\label{uintro}
\end{equation}

This equation describes the evolution of the average synaptic current $u(x,t)$ at spatial location $x$ of the continuous neural tissue and time $t$. The membrane time constant of around 10 ms, is often not explicitly written. $I_\text{ext}(x,t) $ represents an external input, $w(X)$ is the synaptic weight distribution with distance $X$, and $f$ is the firing rate function. This term represents  the effective mean firing rate as a function of an increasing input received by a neuron. Theoretical studies on stochastic conductance-based models have shown that $f$ could typically be described as a sigmoid function~\citep{izhikevichDynamicalSystemsNeuroscience2006}:

\begin{equation}
f(u-\kappa) = \frac{1}{1+e^{-\beta_f(u-\kappa)}}
\end{equation} 
where $\beta_f$ is the gain and $\kappa$ is the threshold.

For a synaptic weight kernel chosen as
\begin{equation}
 w(X)=\left(1-\frac{|X|}{\sigma}\right)e^{-\frac{|X|}{\sigma}},
\label{w}
\end{equation}

This kernel describes a local excitation, and lateral inhibition connectivity. This organization has been found in orientation tuning in the visual cortex~\citep{ben-yishaiTheoryOrientationTuning1995}, and has been used early on in neural fields studies~\citep{amariDynamicsPatternFormation1977}. In particular, its use to model encoding of visual stimuli and working memory makes it suitable for our modelling purposes, see the introduction section. It is however a phenomenological choice, and its application to the SCT processes we aim to describe here would need to be verified.\\
eq.\eqref{uintro} admits a stationary bump solution~\citep{amariDynamicsPatternFormation1977} $u_0(x)$, that can be derived at the limit of infinite gains, where the firing rate function converges to a heaviside function (i.e. $f(u-\kappa)=\Theta(u-\kappa)$ when $\beta_f\xrightarrow{}\infty)$:

\begin{equation}
u_0(x) = \int_{-a}^{a}dy w(x-y) = W(x+a)- W(x-a)
\end{equation}
with $W(X) = Xe^{-\frac{|X|}{\sigma}}$. The bump width, $2a$, is determined considering the boundary conditions: $u_0(\pm a)=\kappa$. 

Equation~\eqref{uintro} constitutes the basis of our model. We extend it here in multiple ways: 

\begin{itemize}
\item We introduce spike frequency adaptation and synaptic depression to modulate the average synaptic current. Both mechanisms have already been implemented and studied in detail in a neural field context in the works of Kilpatrick and Bressloff~\citep{kilpatrickStabilityBumpsPiecewise2010,kilpatrickEffectsSynapticDepression2010}. Spike frequency adaptation describes the attenuation of firing rate after a prolonged period of firing, often due to a calcium-activated potassium current. This process can be implemented as an activity-dependent increase of the threshold $\kappa$ of eq.\eqref{uintro}~\citep{coombesBumpsBreathersWaves2005}. Synaptic depression is related to presynaptic resource depletion. This process can be accounted for in the neural fields equation by introducing an activity-dependent factor $q$ in the nonlocal, integral term of eq.\eqref{uintro}~\citep{kilpatrickStabilityBumpsPiecewise2010,kilpatrickEffectsSynapticDepression2010}.

\item We add long-term synaptic plasticity by introducing a Hebbian plasticity expression in the kernel $w$ of eq.~\eqref{w}. A core hypothesis of our model is that the differences of learning rates between the neocortex and the hippocampus are related to the distance separating distinct elements of an engram, here the bumps of the multiple bump solution. Thus, our Hebbian plasticity rule is modulated by the distance between neurons. In addition to this distance-dependent learning rule, our plasticity equation also includes a slow decay term, that accounts for the degradation of memories in the absence of reactivation.

\item We add a feedback mechanism for the threshold  $\kappa$ that accounts for neurogenesis. As evoked above, the insertion of highly excitable newborn neurons can lead to a decrease of the intrinsic excitability of mature pre-existing neurons. We implement this process by a specific activity-dependent regulation of the threshold $\kappa$.

\item Another particularity of our model is the inter-connection of three individual neural fields, that represent the major regions of the brain involved in the interactions between the hippocampus and the neocortex. These three regions are classically described as part of a particular structure, the trisynaptic pathway~\citep{basuCorticohippocampalCircuitSynaptic2015}, see figure~\ref{intro}~(b), that connects the neocortex (C), to the hippocampus split here into the dentate gyrus (D) and the CA1-CA3 subfields (H), merged here for the sake of simplicity. Hence, our model features three neural fields. In each neural field, the kernel $w$ of eq.~\eqref{w} accounts both for synaptic connections within the neural field and between the two others, in agreement with the trisynaptic pathway. Beyond biological realism related to the trisynaptic pathway, we found that it was necessary to distinguish two separate hippocampal neural fields in the model. Indeed, with a single hippocampal neural field, the highly excitable newborn neurons affect not only the hippocampus but also the neocortical neural field, which erases memory in the neocortex too and prevents retrieval.

\end{itemize}

\section{Methods}
\label{methods}
Our model consists of three connected 1$d$ neural fields, each one representing a brain area. The different variables are indexed with $\alpha$ to refer to the area/neural field they describe, the neocortex ($\alpha=C$), dentate gyrus ($\alpha=D$) and CA regions ($\alpha=H$). In each area, the memory engram takes the form of a double bump attractor pattern, centered around positions $Z^\alpha = \{A^\alpha, B^\alpha\}$, with widths $2a$ (see the ``External stimulation phases'' subsection for further details).\\ Our model is given by the following differential equations.

\paragraph{Activity}
The evolution of activity $u^\alpha(x,t)$ in region / neural field $\alpha$, at position $x$ and time $t$, is described by
\begin{equation}
\frac{\partial u^\alpha}{\partial t}(x,t)=-u^\alpha(x,t)+ I^\alpha(x,t) + I_\text{ext}^\alpha(x,t), 
\label{eq:udiff}
\end{equation}

where $I_\text{ext}^\alpha(x,t)$ corresponds to an external input and $I^\alpha(x,t)$ is the average synaptic current.

\paragraph{Average synaptic current}
$I^\alpha(x,t)$ integrates the neural population activity, averaged over the domain of integration $\Gamma^\alpha$, that comprises the activity of the considered neural field $\alpha$ and that of the neural fields $\beta$ to which it is connected: 

\begin{equation}
I^\alpha(x,t)=\sum_{\beta\in E}G^{\alpha\beta}\int_{\Gamma^\beta}dy q^\beta(y,t)w_\text{tot}^{\alpha\beta}(x,y,t) f(u^\beta(y,t)-\kappa_\text{tot}^\beta(y,t))
\label{eq:Ialpha}
\end{equation}
where $E=\{C,D,H\}$ is the set of field indexes. The constants $G^{\alpha\beta}$ are the elements of the adjacency matrix that defines the directed connectivity of the trisynaptic circuit (fig.~\ref{intro}~(b)). $G^{\alpha\beta}>0$ means that the $\beta$ field sends connections the $\alpha$ field while $G^{\alpha\beta}=0$ means there is no connection from $\beta$ to $\alpha$. Moreover, to equilibrate the total currents received by each field, we fix $G^{\alpha\beta}=1$ for fields $\alpha$ which receive connection from a unique field $\beta$, while for other connections $G^{\alpha\beta}=0.5$, see table~\ref{tab:table-name}.\\
In eq.~\eqref{eq:Ialpha}, $\kappa_\text{tot}^\alpha(x,t)$ is the adaptive threshold, $q^\alpha(x,t)$ is the synaptic scaling factor and $w_\text{tot}^{\alpha\beta}(x,y,t)$ describes the synaptic weights. The activity-dependent expressions of those three terms are given below.

%\begin{equation}
%I_{s}^\alpha(x,t)=\sum_{\beta\in E}G^{\alpha\beta}\eta_s^{\alpha\beta}(x)\int_{\Gamma^\beta}dy q^\beta(y,t) \gamma^{\alpha\beta} s^{\alpha\beta}(x,y,t) f(u^\beta(y,t)-\kappa^\beta(y,t))
%\end{equation}
%\begin{equation}
%\tau_{a_w}\frac{\partial a_w^\alpha}{\partial t}(x,t) = -\bigl(a_w^\alpha(x,t)-[I^\alpha_f(x,t) - \kappa^\alpha(x,t)]_+\bigr)f(u^\alpha(x,t)-\kappa^\alpha(x,t))f(u^\beta(x,t)-\kappa^\beta(x,t))
%\end{equation}
%\begin{equation}
%\eta^{\alpha\beta}_w(x,y,t) = 
%\biggl[I^\alpha_f(x,t) - \bigl[\kappa_{in}+\eta_\kappa f(u^\beta(y,t)-\kappa^\beta(y,t))\bigr]\biggr] 
%\end{equation}

\paragraph{Adaptive thresholds}
\label{thresholdP}
The threshold $\kappa_\text{tot}^\alpha(x,t)$ is composed of two parts:
\begin{equation}
\kappa_\text{tot}^\alpha(x,t) = \kappa^\alpha(x,t) + \kappa_{n}^\alpha(x,t).
\end{equation}
$\kappa^\alpha(x,t)$ implements spike-frequency adaptation while $\kappa_{n}^\alpha(x,t)$ accounts for the effects of neurogenesis.

For spike-frequency adaptation, we mostly follow ref.~\citep{kilpatrickStabilityBumpsPiecewise2010,kilpatrickEffectsSynapticDepression2010,coombesBumpsBreathersWaves2005}: 
\begin{equation}
\tau_\kappa\frac{\partial \kappa^\alpha}{\partial t}(x,t) = -(\kappa^\alpha(x,t)-\kappa_\text{in})+\eta_\kappa f_a(u^\alpha(x,t)-\kappa_\text{in})
\label{eq:kdiff}
\end{equation}
In the original studies of this equation, the baseline threshold in the first term of the right-hand side and the threshold of the firing function $f_a$ of the second term have different values~\citep{coombesBumpsBreathersWaves2005,kilpatrickStabilityBumpsPiecewise2010,kilpatrickEffectsSynapticDepression2010}.
However, detailed biophysical models suggest that it is more biologically realistic to use the same parameter in the two terms, here $\kappa_\text{in}$~\citep{bendaUniversalModelSpikeFrequency2003}. $f_a$ is a firing function smoother than $f(u-\kappa) = 1/\left(1+e^{-\beta_{f}(u-\kappa)}\right)$ in eq.~\eqref{eq:Ialpha}, i.e. $f_a(u-\kappa) = 1/\left(1+e^{-\beta_{f_a}(u-\kappa)}\right)$ with $\beta_{f_a}<\beta_f$ (see below for a justification). We refer to the smoothness of such sigmoid functions as the small derivative of their slope. The characteristic time scale of the process $\tau_\kappa$ has been experimentally found to lie between $40$ and $120$ ms~\citep{madisonControlRepetitiveDischarge1984}.

The neurogenesis term $\kappa_{n}^\alpha(x,t)$ is expected to emulate the progressive adaptation mechanism responsible for the local decrease of excitation in reaction to the integration of highly excitable newborn DG neurons. For simplicity, we use the same evolution equation as for spike frequency adaptation, however with a much slower timescale ($\tau_{\kappa_n}>>\tau_{\kappa}$) and a lower adaptation threshold ($\theta_n<\kappa_\text{in}$):

\begin{equation}
\tau_{\kappa_n}\frac{\partial \kappa_n^\alpha}{\partial t}(x,t) = -\kappa_n^\alpha(x,t)+\eta_{\kappa_n} \Theta(u^\alpha(x,t)-\theta_n)
\label{eq:kndiff}
\end{equation}
Here we use a heaviside function $\Theta$ for simplicity instead of the smoother firing rate function of eq.~\eqref{eq:kdiff}. We leave for future works the evaluation of the impact of this simplification.

To emulate the high excitability of newborn D neurons, the threshold of the neurons located around the areas involved in the engram is lowered to a value $g_n \kappa_\text{in}$ with $g_n < 1$ in eq.~\eqref{eq:kdiff}. We only consider neighbour neurons because more distant neurons would have a very small effect. The spatial domain of these newborn neurons is therefore

\begin{equation*}
\bigcup_{z\in Z^D}\biggl([z-a-\delta_{n},z-a]\cup[z+a,z+a+\delta_{n}]\biggr)
\end{equation*}
with $\delta_{n}$ the width of newborn neurons subarea. \\

\paragraph{Synaptic scaling}
The evolution of the synaptic scaling factor $q^\alpha(x,t)$ emulates synaptic depression. Here again we mostly follow the implementation proposed by Kilpatrick and Bressloff (2010)~\citep{kilpatrickStabilityBumpsPiecewise2010,kilpatrickEffectsSynapticDepression2010}:

\begin{equation}
\frac{\partial q^\alpha}{\partial t}(x,t) = \frac{1-q^\alpha(x,t)}{\alpha_q} -m_q^\alpha(x,t) q^\alpha(x,t) f_a(u^\alpha(x,t)-\kappa_\text{tot}^\alpha(x,t))
\label{eq:qdiff}
\end{equation}

In contrast to the original equation however, we introduce an activity dependent term for synaptic depletion that allows to keep bump durations roughly constant when learning weights evolve and/or neurogenesis is applied:

\begin{equation}
m_q^\alpha(x,t) = \beta_q u^\alpha(x,t)
\label{eq:mq}
\end{equation}

% The term acts as a switch, see Figure~\ref{K}. $K^\alpha$ does not affect the dynamics until it reaches a critical value which leads the threshold $\kappa^\alpha$ to exceed the activity $u^\alpha$ inside the bump, thus stopping the firing.

\paragraph{Synaptic weights and plasticity}
The kernel $w^{\alpha\beta}_\text{tot}(x,y,t)$ of eq.~\ref{eq:Ialpha} is composed of two terms:
\begin{equation}
w_\text{tot}^{\alpha\beta}(x,y,t)=w(x-\Delta^{\alpha\beta}(x)-y)+\gamma^{\alpha\beta}\eta_s^{\alpha\beta}(x,t)s^{\alpha\beta}(x,y,t)
\label{eq:wtot}
\end{equation}
\begin{itemize}
\item $w(X)$ is given by eq.~\eqref{w} above. It represents a permanent (learning-independent) connectivity between a post-synaptic neuron at position $x$ and a presynaptic neuron at position $y$. However, compared to the expression used in classical neural field formulations, we introduce a position-dependent shift $\Delta^{\alpha\beta}(x) = (B^\alpha-B^\beta)\sgn(x)=-(A^\alpha-A^\beta)\sgn(x)$. This shift emulates the difference in size of the neocortical and hippocampal fields. Using this trick, despite the intrinsic difference of distances between pattern locations ($|A^C-B^C|>|A^{D}-B^{D}|$ and $|A^{D}-B^{D}|=|A^{H}-B^{H}|$), $\Delta^{\alpha\beta}(x)$ allows to force the permanent connection between $A^C$ and $A^H$ (or $A^D$), $B^C$ and $B^H$ (or $B^D$). For instance, for $x=A^C$,\\ $w(A^C-\Delta^{CH}(A^C)-A^H) = w(A^C-\Delta^{CC}(A^C)-A^C) = w(0) = 1$, i.e. the weight between $A^C$ and $A^C$ is the same as between $A^C$ and $A^H$.\\ This phenomenological description of inter-field interactions for bumps separated by different distances is a highly simplified modelling choice. It is here made possible, as there are only two bumps at defined positions in each field. In a case where bumps could be anywhere on the fields, this space shift would not be a satisfying description, and it would be more suitable to implement directly a larger neocortical field.

\item Synaptic plasticity rules in neural field models can take the form of an immediate modulation of the kernel depending on the neuronal activities~\citep{abbassianNeuralFieldsFast2012}. However in the case of long-lasting synaptic modifications it is interesting to introduce an additional differential equation with a larger characteristic timescale. A rate-based plasticity rule preserving temporal correlations has been introduced for connections between pre- and post-synaptic neural fields~\citep{robinsonNeuralFieldTheory2011}, and has been used in particular for topographic maps~\citep{galeAnalysisActivityDependent2022,detorakisNeuralFieldModel2012}. However such temporal correlations are not important for our model, and we preferred a simpler Hebbian expression that preserves mathematical and computational tractability. Therefore, we implement here synaptic plasticity by the second term of eq.~\eqref{eq:wtot}: $\gamma^{\alpha\beta}\eta_s^{\alpha\beta}(x,t)s^{\alpha\beta}(x,y,t)$, where the $\gamma^{\alpha\beta}$s are positive constants, assumed to be larger for intra than for inter-field connections, $\gamma^{\alpha\alpha}>\gamma^{\alpha\beta}$ for $\alpha\neq \beta$, to represent stronger interactions within one brain area.\\
$s^{\alpha\beta}(x,y,t)$ is a variable that represents the learning weight. Its dynamics is given by two additive terms:
\begin{equation}
\frac{\partial s^{\alpha\beta}}{\partial t}(x,y,t) = L^{\alpha\beta}(x,y,t) - F^{\alpha\beta}(x,y,t)
\label{eq:sdiff}
\end{equation}
The first term $L^{\alpha\beta}(x,y,t)$ is a bounded, distance-dependent, Hebbian~\citep{gerstnerMathematicalFormulationsHebbian2002} learning rule:
\begin{equation}
L^{\alpha\beta}(x,y,t) = (1-s^{\alpha\beta}(x,y,t))d(x,y)\times f(u^{\alpha}(x,t)-\kappa_\text{tot}^{\alpha}(x,t))f(u^{\beta}(y,t)-\kappa_\text{tot}^{\beta}(y,t)),
\label{eq:LweightsL}
\end{equation}
where distance modulates the learning rate via the factor:
\begin{equation}
 d(x,y)=\frac{A_d}{\sigma_d}e^{-|x-y|/\sigma_d}.
\label{eq:dLR}
\end{equation}
With eq.~\eqref{eq:dLR}, closer $x$ and $y$ positions yield larger $d(x,y)$, thus faster learning rates. The second term $F^{\alpha\beta}$ is a forgetting, decay term, which occurs much more slowly than the first term ($c_0<<1$).
 
\begin{equation}
F^{\alpha\beta}(x,y,t) = 
\begin{cases}
\frac{c_0}{s^{\alpha\beta}(x,y,t)} & \text{ if } s^{\alpha\beta}(x,y,t)>0\\
0 & \text{ if } s^{\alpha\beta}(x,y,t) = 0
\end{cases}
\label{eq:LweightsS}
\end{equation}

Introducing the  $s^{\alpha\beta}(x,y,t)$ term in the right-hand side of eq.~\eqref{eq:wtot} generates a discontinuity in the activity~\citep{fotouhiContinuousNeuralNetwork2015}. To maintain the continuity in $u^{\alpha\beta}(x,t)$ expression, we therefore add a continuity function $\eta_s^{\alpha\beta}(x,t)$ in eq.\eqref{eq:wtot} : 

\begin{equation}
\eta_s^{\alpha\beta}(x,t) = 
\Biggl[\int_{\Gamma^\beta}dy w(x-\Delta^{\alpha\beta}(x)-y)\Theta(s^{\alpha\beta}(x,y,t)) - \kappa_\text{in}\Biggr] _+
\label{eq:etas}
\end{equation}
with the rectification function $[v]_+ = v \text{ if } v>0, \ 0 \text{ otherwise}$.

When $s^{\alpha\beta}(x,y,t)>0$, i.e. for neurons that are part of the memory pattern, the integral in eq.\eqref{eq:etas} corresponds to the received bump current of one field, which is $\kappa_\text{in}$ at boundaries. This guarantees continuity at boundaries, see the supplementary materials section~\ref{appendices} for more details.

\end{itemize}

\begin{figure}
\centering
\includegraphics[height=6.5cm]{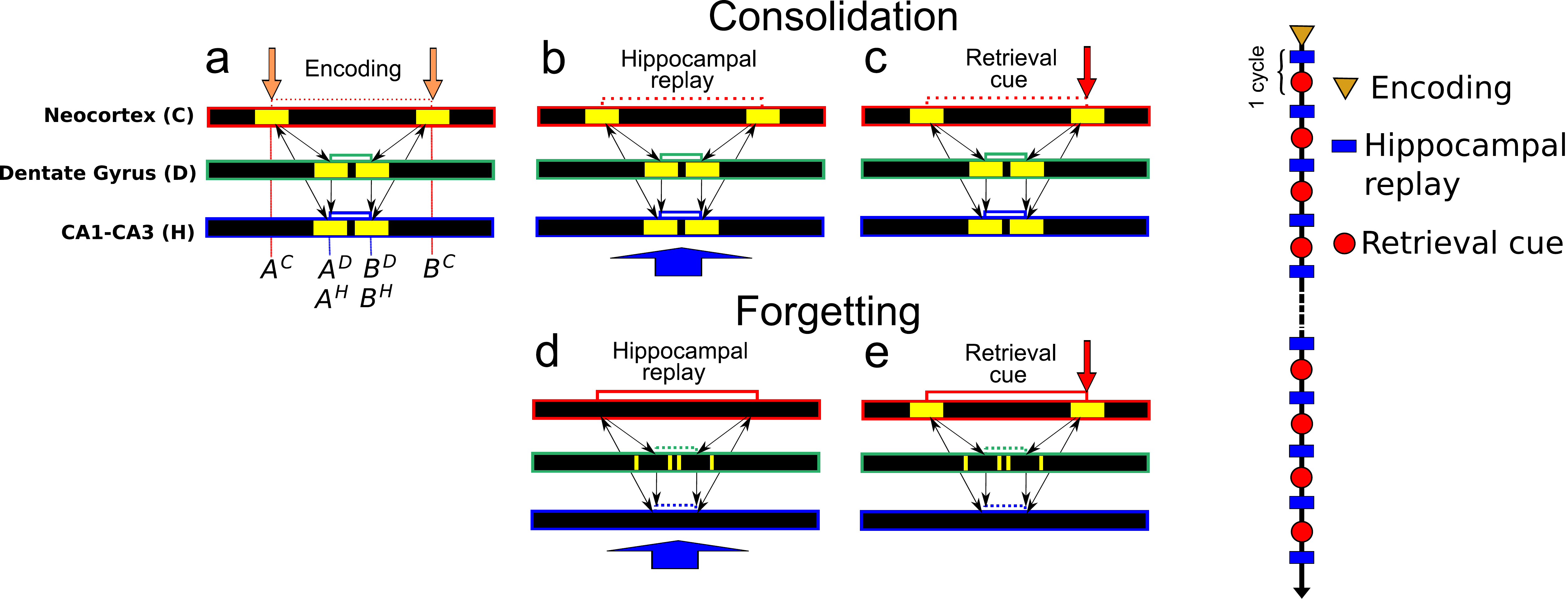}
\caption[Temporal sequence describing standard consolidation theory key moments]{Temporal sequence describing the consolidation process according to standard consolidation theory. \textbf{(Left)} Different steps of the computational model. \textbf{(a)} The first diagram represents the encoding phase, where two external signals arrive on the neocortical field at two distinct positions, $A^C$ and $B^C$. The activated neocortical neurons that now encode the memory pattern fire, which activates other neurons in the two other hippocampal fields (D, H), at positions $A^{D}$ and $B^{D}$ (and $A^{H}$ and $B^{H}$). 
After the initial encoding, cycles of hippocampal replay and retrieval cue steps are repeated, which allows consolidation. \textbf{(b)} The hippocampal replay step occurs during sleep. We model it by a spatially uniform external stimulation on the CA region (H). 
\textbf{(c)} The retrieval cue is a partial external input centered on only $A^C$ or $B^C$ in the neocortex. \textbf{(d,e)} Following experimental suggestions~\citep{franklandHippocampalNeurogenesisForgetting2013}, we model the effect of neurogenesis in the dentate gyrus (D) by reducing the threshold of newborn neurons located in the neighborhood of the pattern (see text). \textbf{(Right)} Summary of the steps for the whole process. After the encoding phase, cycles of hippocampal replay followed by retrieval cue step are iterated. The time between two steps there is a long waiting period, 100 times longer than the duration of a pattern reactivation.}
\label{mainsteps}
\end{figure}

\paragraph{External stimulation phases}
\label{ext}
To emulate the repetition of reactivations of hippocampal memory patterns and presentations of the pattern to the neocortex that are central to the SCT,  we submitted the model to a sequence composed of hundreds of alternation of two phases, see figure~\ref{mainsteps}. The initial encoding is simulated by the application of an external input on the neocortex that represents the experience to be memorised. This external input can be encoded as a memory pattern in the three neural fields in the form of a double bump attractor pattern. At this initial encoding stage, the external current is non-zero only in the neocortex (C), where it assumes a rectangular function, located around positions $A^C$ and $B^C$,  with widths equal to $2a$. These characteristics define the information contained in the memory pattern. The external input to the neocortex during the initial encoding stage thus is: 
\begin{equation}
I_\text{ext}^C(x,t) = G^{C\text{Ext}}\sum_{z\in Z^C} \biggl[\Theta(x-(z-a))-\Theta(x-(z+a))\biggr] i^C(t)
\label{eq:IextC}
\end{equation}
where $i^C(t)$ is an indicator function equals to $1$ during the encoding phase, $0$ otherwise. After this initial encoding, we iterate cycles composed of a hippocampal replay (HR)-- assumed to occur offline, during sleep-- followed by a retrieval cue (RC) -- assumed to occur online, during a waking period. In this sequence, one cycle corresponds approximately to one day. 

HR steps are simulated in the model by the application of an homogeneous external input on the CA region (H), that is assumed to emulate slow-wave sleep~\citep{klinzingMechanismsSystemsMemory2019}. Hence, during HR steps, the external current targets the H region only, with $I_\text{ext}^H(x,t)$ uniform over the neural field space, $I_\text{ext}^H(x,t)=G^{H \text{Ext}},\  \forall x$. An RC step corresponds to the application of a partial external input on the neocortex only, that stimulates only one of the two bumps of the encoded pattern. Stimulation of a single bump represents an experience partially similar to the encoded pattern (that exhibits 2 bumps). Memory retrieval is achieved if the stimulation of this single bump activates the totality of the 2-bump encoded memory. Therefore, the external current during RC steps assumes a form similar to eq.~\eqref{eq:IextC}, except that $Z^C$ is restricted to $A^C$ or $B^C$ (in alternation). In every steps, the external input is maintained for a short period of time (see table~\ref{tab:table-name}).

\paragraph{Numerical simulations}
We solved the model eq.~\eqref{eq:udiff} to~\eqref{eq:etas} on a one-dimensional spatial domain with periodic boundaries, that is, on ring of length total length $x_{m1}-x_{m2}=60$ centered at $x=0$. The equations were solved using a forward explicit Euler method with spatial increment $dx= 0.08$ and time increment $dt = 0.1$. Numerical integration of the integral was carried out as a sum of rectangles of width $dx$, times $dx$.\\
The code used to generate the results of this article is freely available online on GitLab: \url{https://gitlab.inria.fr/lisa.blum-moyse/a-coupled-neural-field-model-for-the-standard-consolidation-theory}. 

\begin{table}
\centering
\begin{tabular}{ |c|c| } 
\hline
Symbol &  Definition \\
\hline
\multicolumn{2}{|c|}{Time varying functions}\\
\hline
$u^\alpha$ & Activity \\
$f$ & Firing function\\
$f_a$ & Firing function for $\kappa^\alpha$ and $q^\alpha$ equations \\
$I^\alpha$ & Average synaptic current\\
$I^\alpha_{ext}$ & External current\\
$s^{\alpha\beta}$ & Learning weight\\
$\eta_s^{\alpha\beta}$ & Continuity function\\
$\kappa_\text{tot}^\alpha$ & Threshold\\
$\kappa^\alpha$ & Threshold - adaptation\\
$\kappa_{n}^\alpha$ & Threshold - neurogenesis\\
$q^\alpha$ & Synaptic scaling factor\\
\hline
\multicolumn{2}{|c|}{Time independent functions}\\
\hline
$w$ & Permanent weight\\
$W$ & Integral of the permanent weight\\
$d$ & Distance function\\
\hline
\end{tabular}
\caption[Main functions used for the neural field model]{\label{tab:functions} Main functions used for the neural field model}
\end{table}

\begin{table}
\centering
\begin{tabular}{ |c|c|c| } 
\hline
Parameter & Value & Definition \\
\hline
\multicolumn{3}{|c|}{Structure and positions}\\
\hline
E & - & Set of field indexes\\
C & - & Neocortex field index\\
D & - & Dentate gyrus field index\\
H & - & CA regions field index\\
$A^C$ & -16 & Position of the C left pattern location \\
$B^C$ & 16 & Position of the C right pattern location \\
$A^D$ & -10 & Position of the D left pattern location \\
$B^D$ & 10 & Position of the D right pattern location \\
$A^H$ & -10 & Position of the H left pattern location \\
$B^H$ & 10 & Position of the H right pattern location \\
a & 0.9 & Bump widths in all fields\\
$G^{CC}$ & 1 & Amplitude of the current from C to C \\
$G^{DD}$ & 1& Amplitude of the current from D to D \\
$G^{HH}$ & 1& Amplitude of the current from H to H \\
$G^{CH}$ & 1& Amplitude of the current from H to C \\
$G^{CD}$ & 0& Amplitude of the current from D to C \\
$G^{HD}$ & 0.5& Amplitude of the current from D to H \\
$G^{HC}$ & 0.5& Amplitude of the current from C to H \\
$G^{DC}$ & 1& Amplitude of the current from C to D \\
$G^{DH}$ & 0& Amplitude of the current from H to D \\
$\sigma$ & 1.5 & Width of the permanent weights\\
\hline
\multicolumn{3}{|c|}{Firing rate}\\
\hline
$\beta_f$ & 250 & Gain of the firing function\\
$\beta_{f_a}$ & 50 & Gain of the firing function for $\kappa^\alpha$ and $q^\alpha$ equations\\
\hline
\multicolumn{3}{|c|}{Learning kernels}\\
\hline
$\gamma$&1.5& Constant for intra-field learning weights ($\gamma^{\alpha\alpha}=\gamma$)\\
$cr$ & 0.2& Factor for the constant for inter-field learning weights ($\gamma^{\alpha\beta}=cr\times\gamma$, $\alpha\neq\beta$)\\
$A_d$ & $3$ & Amplitude of the distance function\\
$\sigma_d$ & $9$ & Width of the distance function \\
$c_0$ & $8.10^{-7}$ & Decay rate\\
%$\epsilon_w$ & $10^{-2}$ & Constant of the decay term\\
\hline
\multicolumn{3}{|c|}{Thresholds}\\
\hline
$\kappa_\text{in}$ & 0.54 & Baseline threshold\\
$\tau_\kappa$ & 0.8 & Time scale for spike frequency adaptation\\
$\eta_\kappa$ & $0.54$ & Strength for spike frequency adaptation \\ 
$\tau_{\kappa_n}$& 1000 & Time scale of the slow adaptation related to neurogenesis\\
$\eta_{\kappa_n}$ & 0.5 & Strength of the slow adaptation related to neurogenesis\\
$\theta_{n}$ & 0.001 & Baseline threshold for slow adaptation related to neurogenesis\\
\hline
\multicolumn{3}{|c|}{Synaptic scaling}\\
\hline
$\alpha_q$ &  800 & Time scale of synaptic resources recovery\\ 
$\beta_q$ & 0.01  & Inverse of the time scale of synaptic resources depletion\\
\hline
\multicolumn{3}{|c|}{External currents}\\
\hline
$G^{C \text{Ext}}$& 1.5 & Amplitude of the external current in C\\
$G^{H \text{Ext}}$& 0.87 & External current value in H\\
$-$& 1.8 & Duration of the external stimulation in C\\
$-$& 1.2 & Duration of the external stimulation in H\\

\hline
\multicolumn{3}{|c|}{Neurogenesis}\\
\hline
$\delta_{n}$ & 0.32 & Width of the newborn neurons area \\
$g_n$& 0.1 & Factor of newborn neurons reduced threshold\\
\hline
\multicolumn{3}{|c|}{Numerical parameters}\\
\hline
$x_{m1}$ & -30 & Left extremity of a field\\
$x_{m2}$ & 30& Right extremity of a field\\
$dx$ & 0.08 & Spatial increment\\
$dt$ & 0.1 & Time increment\\
\hline
\end{tabular}
\caption[Parameters values used for the neural field model]{\label{tab:table-name} Parameters values used for the neural field model. Because of the abstract nature of the model, units were not taken into account here. See the section~\ref{discussion} for a discussion on the timescale parameters.}
\end{table}

\section{Results}
\label{ResultsNF}
As illustrated in fig.~\ref{mainsteps}, we submitted our coupled neural field model, defined by eq.~\eqref{eq:udiff} to~\eqref{eq:etas} to cycles of repeated external inputs. We first study below the consolidation mechanism, then the forgetting effect of neurogenesis is tested on consolidated memory patterns. Figure~\ref{mainsteps} is used as a visual support throughout this results section and as a guide to the different phases of the process.

For convenience, throughout the paper we refer to neurons located within the excited region around $A^\alpha$ ($[A^\alpha-a,A^\alpha+a]$) or $B^\alpha$ ($[B^\alpha-a,B^\alpha+a]$), as respectively $A^\alpha$ or $B^\alpha$ neurons. Furthermore, to monitor the temporal evolution of synaptic learning along the repetition cycles, we show the evolution of the learning weights $s^{\alpha\beta}(x,y,t)$ instead of the total weights $w_\text{tot}^{\alpha\beta}(x,y,t)$, eq.~\eqref{eq:wtot}, because they constitute a relevant and concise measure of synaptic plasticity, since they are bounded between $0$ and $1$.  

\subsection{Consolidation}
\label{consolidationP}
\begin{figure}
\centering
\includegraphics[width=13cm]{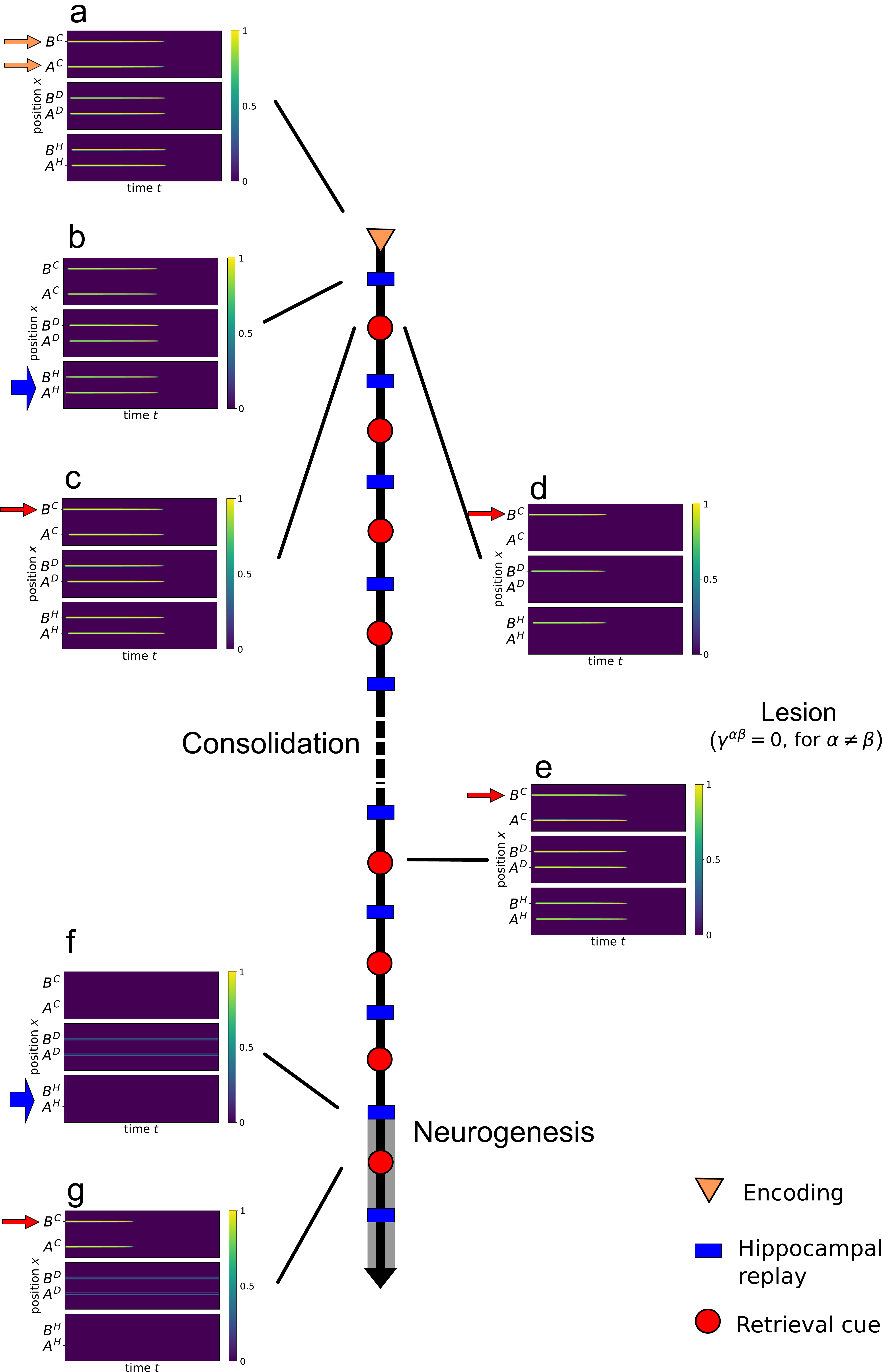}
\caption[Snapshots in the temporal sequence of firing rates in the three fields]{Snapshots of firing rates in the three fields, neocortex (C), dentate gyrus (D) and CA regions (H) along the repetitions of the temporal sequence . \textbf{(a)} The encoding step shows a first pattern activation in the three fields. The initial stimuli around positions $A^C$ and $B^C$ in the neocortical field activate C neurons which in turn activate D and H neurons, yielding bump attractors in all neural fields. \textbf{(b)} The hippocampal replay step allows engram reactivation. We model hippocampal reactivation by sending an uniform stimulus over space on neural field H. \textbf{(c)} The retrieval cue step allows engrams reactivation. This step consists in stimulating only the $A^C$ or $B^C$ location of the neocortex (here $B^C$). Retrieval is exhibited by the subsequent activation of $A^C$. \textbf{(d,e)} A ``lesion'' between the hippocampal and neocortical neural fields is simulated by setting $\gamma^{\alpha\beta}=0$ for all $\alpha \neq \beta$ in eq.\eqref{eq:wtot}. \textbf{(d)} At early stages, the activation of the $B^C$ neurons during retrieval fails to activate the $A^C$ neurons if hippocampal learning is not functional. \textbf{(e)} However, at the end of the consolidation process, pattern retrieval in the neocortex is possible even in the absence of a functional hippocampus. \textbf{(f,g)} HR and RC when neurogenesis is effective. The thin lines in the D field evidence the continuous firing of those new neurons. \textbf{(f)} Neurogenesis prevents the reactivation of hippocampal engrams during the HR phase because of the inceased thresholds caused by the excitable newborn D neurons. \textbf{(g)} Neurogenesis also prevents the reactivation of hippocampal engrams during neocortical RC.}
\label{AllSCT}
\end{figure}

\paragraph{Encoding}
The process is initiated by the encoding step, where an external signal stimulates two distinct areas around the positions $A^C$ and $B^C$, in the neocortex (C), see the schematized process in figure~\ref{mainsteps}~(a) and the corresponding numerical simulation in figure~\ref{AllSCT}~(a). The stimulated cortical (C) neurons then induce firing in the dentate gyrus (D), and, finally, in the CA regions (H) where the H neurons receive inputs from C and D. The interconnection of these neural fields which are quickly activated one after the other modifies the total currents in the activity equation eq.\eqref{eq:udiff}. This variation takes place quickly, shortly after the external input. Since spike frequency adaptation is active at this timescale, the firing threshold increases from its initial value $\kappa_\text{in}$ to a value $\kappa_\text{in}+\eta_\kappa$. This term is responsible for the disappearance of the bumps after a time (fig.~\ref{AllSCT}~(a)). Spike frequency adaptation also maintains the bump width, which characterizes the information of the memory pattern (see section~\ref{appendices} for the relation between these threshold values and their width). However, numerical exploration of the model showed that adaptation was not able to properly and quickly enough adjust to preserve the bump width, unless its firing function ($f_a$ in eq.~\eqref{eq:kdiff}) is smoother than the firing function $f$ of the activity equation  eq.\eqref{eq:udiff}, i.e. $\beta_{f_a}<\beta_f$.

\begin{figure}
\centering
\includegraphics[height=5cm]{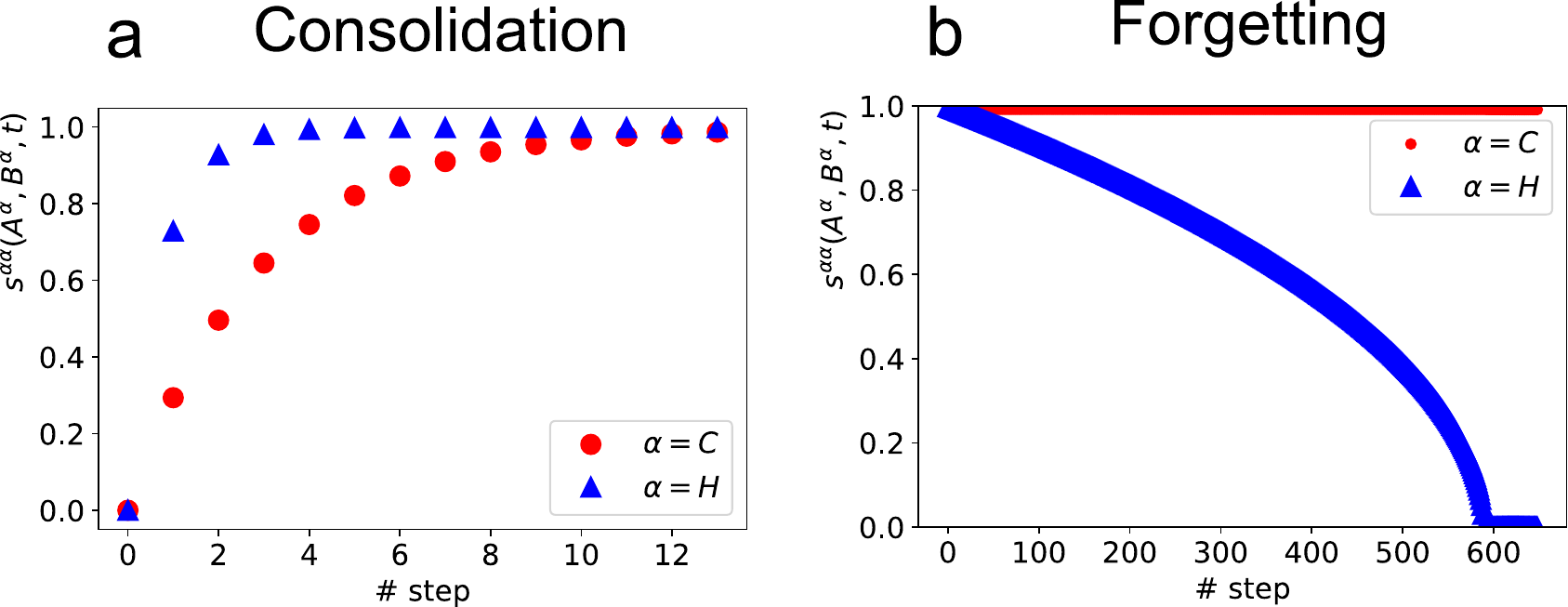}
\caption[The learning weights $s^{CC}(A^C,B^C,t)$, $s^{HH}(A^H,B^H,t)$ evolution in the neural fields model highlights the slow learning, stable memory features of the neocortex and the fast learning, unstable memory features of the hippocampus]{Evolution of the learning weights $s^{CC}(A^C,B^C,t)$ (\textit{red circles}) and $s^{HH}(A^H,B^H,t)$ (\textit{blue triangle}) at the beginning of every consolidation step. $s^{DD}(A^D,B^D,t)$ displays the same dynamics as $s^{HH}(A^H,B^H,t)$. \textbf({a}) During the consolidation phase, the $A^H-B^H$ weights rise faster than the $A^C-B^C$ ones: the intra-hippocampal weights reach the maximal value $1$ after approximately $4$ steps, while this is achieved in $12$ steps for neocortical ones. \textbf({b}) When neurogenesis is effective, the hippocampal pattern cannot be retrieved anymore (fig.~\ref{AllSCT}~(f)~(g)), so that $s^{HH}(A^H,B^H,t)$ slowly decreases. On the opposite, since the whole neocortical pattern is retrieved independently of the hippocampal fields (fig.~\ref{AllSCT}~(g)), $s^{CC}(A^C,B^C,t)$ remain  at the maximum value $1$.}
\label{Sevol}
\end{figure}

As long as the bump pattern is present, the firing of the corresponding neurons leads to an increase of the learning weights through eq.~\eqref{eq:LweightsL}. Therefore, during the encoding phase, all the weights increase between neurons located in an engram location ($[A^\alpha-a,A^\alpha+a]$ and $[B^\alpha-a,B^\alpha+a]$) because of Hebbian synaptic plasticity (see eq.\eqref{eq:LweightsL}). However as illustrated in figure~\ref{Sevol}a, the synaptic weights between the neocortical neurons $A^C$ and $B^C$ increase much more slowly than the intra-hippocampal weights $A^H-B^H$. This difference in learning rate is due to the distance-modulated learning rate of eq.~\eqref{eq:dLR}. Especially since $|A^C-B^C|>|A^D-B^D|$ ($|A^H-B^H|=|A^D-B^D| $), we get 
$d(A^D-B^D)>d(A^{C}-B^{C})$ in  eq.~\eqref{eq:dLR}. In other words, consolidation is slower in the neocortex (C) because the distance between the two bumps of memory pattern is larger than in the hippocampus. Therefore, at the end of the encoding phase, the synaptic weights between the two bumps of the memory pattern in the neocortex are much smaller than those between the two parts of the hippocampal patterns, or between the neocortex and the hippocampus (see figure~\ref{encoding} in the supplementary material).

\paragraph{Hippocampal replay} Repeated reactivations of the memory pattern during sleep, which originates in the hippocampus, consolidate the cortical encoding is through see the schematized hippocampal replays in figure~\ref{mainsteps}~(b) and the numerical simulations in figure~\ref{AllSCT}~(b). During hippocampal replay, an external signal stimulates all neurons of the CA regions (H) neural field. However, only those neurons involved in the pattern locations fire for a long time. Indeed, the current they receive has a higher value, since their learning weights are stronger. This firing in H leads to pattern reactivations in D and C, see fig.~\ref{AllSCT}~(b). The two parts of the memory pattern being active, the weights between $A$ and $B$ neurons in each neural field grow. When the HR is repeated for a sufficient number of times, this also increases the intra-cortical weights between $A^C$ and $B^C$, as shown in fig.~\ref{Sevol}a, thus stabilising the cortical engram.

\paragraph{Retrieval cue}
Since the learning weights between $A^\alpha$ and $B^\alpha$ grow faster in the hippocampus, neocortical memory in the early steps of the process can be retrieved only with the hippocampal neurons, as illustrated in the numerical simulation of figure~\ref{AllSCT}~(c). During this step, only the one bump location is stimulated, $A^C$ or $B^C$, in alternation. Retrieval is achieved when this stimulation of the stimulated bump, e.g., $B^C$ leads to the activation of non-stimulated one (here, $A^C$). However at the initial stages of the process, the activation of $A^C$ cannot happen directly from $B^C$ to $A^C$, it goes via the hippocampus. Indeed, at these initial stages, the weights between $A^C$ and $B^C$ are still small, the activation of the $B^C$ neurons during retrieval first activates the $B^D$ neurons; the $B^H$ neurons are then activated by $B^C$ and $B^D$, which activates the $A^H$ neurons because the $B^H-A^H$ learning weights are quickly large at the initial stages of the process (fig.~\ref{Sevol}a). Finally, it is the $A^H$ neurons that activate the $A^C$ neurons via the strong $A^C-A^H$ weights, leading to the whole pattern recovery via the complete reactivation of $A^C$, see figure~\ref{AllSCT}~(c). Here again, recovering the activity in the two bumps of the neocortical memory pattern ($[A^C-a,A^C+a]$ and $[B^C-a,B^C+a]$), contributes to progressively increase the intra-cortical weights $A^C-B^C$ and thus consolidate the neocortical engram, see figure~\ref{Sevol}~(a) and supplementary figure~\ref{RC}~(b)(c).

To illustrate further that the neocortex is dependent on the hippocampal neural fields for its engram retrieval in the initial stages of consolidation, we performed the simulation shown in fig.~\ref{Sevol}a-c of a simulated hippocampal lesion. Deletion of the whole connectivity $w^{\alpha\beta}_\text{tot}$ stops transmission between the fields, and forbids any retrieval in the hippocampal fields in the retrieval cue step for example. We thus choose here to study the case of a “partial” lesion, keeping the permanent connectivity intact. The hippocampal lesion was simulated by cancelling the learning weights in the hippocampal fields, between them and with the neocortical field, that is, we set $\gamma^{\alpha\beta}=0$ for $\alpha\neq \beta$ in eq.\eqref{eq:wtot}. In this configuration, the complete neocortical pattern cannot be retrieved: activating $B^C$ neocortical neurons now fails to activate $A^C$ neurons (figure~\ref{AllSCT}~(d)). Similarly, the weights between $A^\alpha$ and $B^\alpha$ neurons fail to consolidate (supplementary figure~\ref{RCtest}~(b)(c)). Hence, during the initial stages of the consolidation process, memory cannot be retrieved without a functional hippocampus in our model.

\paragraph{End of consolidation}
After approximately 6 cycles (equivalent to six days) of RC and HR steps, the learning weights are fully consolidated everywhere, see figure~\ref{Sevol}~(a). In particular, strong weights now connect $A^C$ and $B^C$ neurons in the neocortex, which was the longest process to establish. From this stage on, simulation of a retrieval cue step under an hippocampal ``lesion'', as presented above, exhibits a complete neocortical pattern retrieval: the activation of $B^C$ neurons is enough to activate the $A^C$ bump, even in the absence of a functional hippocampal learning, see figure~\ref{AllSCT}~(e).
Hence, in opposition to the initial stages of the consolidation process where neocortical pattern retrieval was not possible without the hippocampus learning weights, retrieval becomes independent of the hippocampal fields at the end of the consolidation process, since the $A^C-B^C$ weights have become strong enough.

\subsection{Forgetting}
\label{forgettingP}
While neocortical memories, once consolidated, can persist for years, the SCT stipulates that hippocampal patterns disappear quickly. To explore this erasure part of the dynamics, we followed the neurogenesis hypothesis (see Methods). We focus on neurons that are located in the neighbourhood of the bump pattern and assume that this region of the neural field is where newborn neurons integrate. In the model, we actually reproduce this process by imposing to this region a tenfold decrease of the firing threshold $\kappa_\text{in}$ in eq.~\eqref{eq:kdiff}. Since these neurons fire continuously for a long time even in the absence of input (see fig.~\ref{AllSCT}~(f)(g)), the slow adaptation term $\kappa_{n}^\alpha(x,t)$ of eq.~\eqref{eq:kndiff} in the neighboring regions of the two hippocampal neural fields (D and H) reaches its stationary value $\eta_{\kappa_n}$. This saturation of the thresholds of the hippocampal pattern neurons prevents the formation of bump attractors when external currents are applied. As a result, with neurogenesis, the HR and the RC steps both fail to evoke the two bumps pattern in the hippocampal neural fields, see figure~\ref{AllSCT}~(f)(g). Neurogenesis thus induces both a reduced excitability of the pattern neurons and a failure to reactive hippocampal memory patterns. The neurons  of the pattern in the hippocampal fields consequently exhibit a long-lasting period of silence: the $L^{\alpha\beta}(x,y,t)$ terms in the equation of their synaptic plasticity (eq.~\eqref{eq:sdiff}) progressively decreases, while the forgetting term $F^{\alpha\beta}(x,y,t)$ grows. After some steps, $s^{\alpha\beta}{\partial t}(x,y,t)$ eventually vanishes, a point at which the hippocampal pattern is effectively erased. 

Therefore, with neurogenesis, hippocampal memory patterns failed to be retrieved, contrary to neocortical ones during retrieval cues since the C memory pattern at this stage of the consolidation process is strong enough to be retrieved entirely. After approximately 300 cycles (equivalent to roughly 300 days) our model exhibits a complete erasure of hippocampal memory pattern while the neocortical one is still strong and retrieved independently of the hippocampus (fig.~\ref{Sevol}~(b) and supplementary figure~\ref{NG}~(c)(d)).

\section{Discussion}
\label{discussion}
\begin{figure}
\centering
\includegraphics[width=12cm]{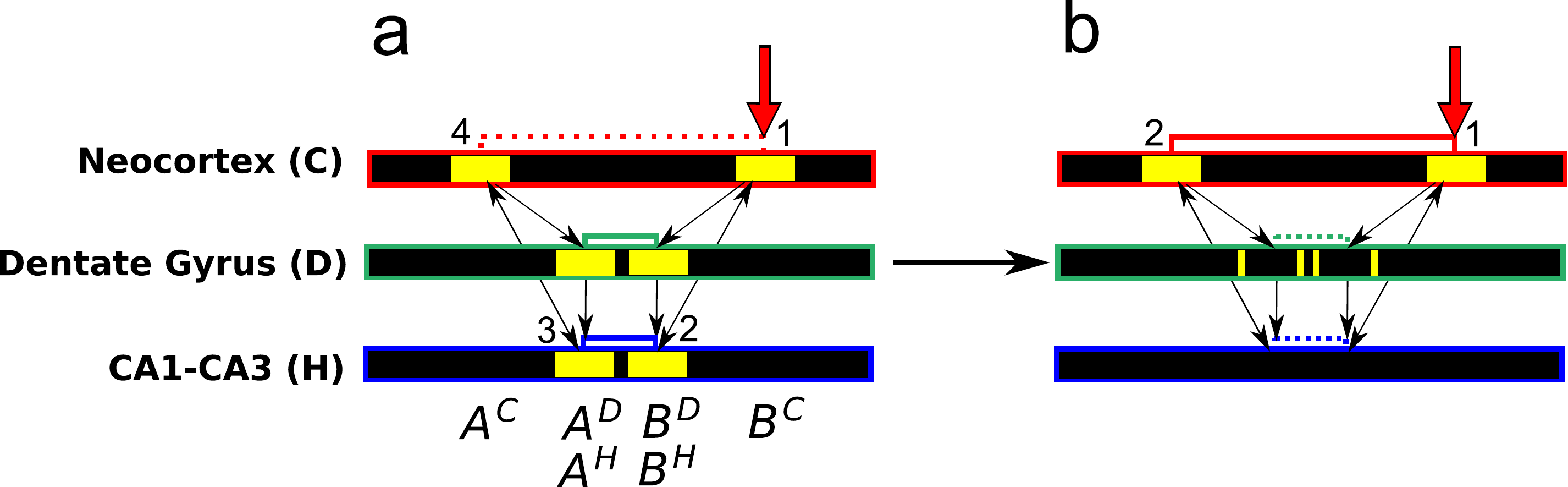}
\caption[Evolution from a dependent to an independent neocortical pattern retrieval in the connected neural fields]{Evolution from a dependent to an independent neocortical pattern retrieval in the connected neural fields. The retrieval cue is a partial signal on only $A^C$ or $B^C$ in the neocortex, here $B^C$ (vertical red arrow). \textbf{(a)} At the beginning of the whole process, when the neocortex is dependent on the hippocampus, the activity is transmitted following $1\xrightarrow{}2\xrightarrow{}3\xrightarrow{}4$. $A^C$ is activated via the hippocampal fields. \textbf{(b)} However at the end of the process, the cortical weights between $A^C$ and $B^C$ are strong enough to activate each other directly ($1\xrightarrow{}2$), independently of the hippocampus, into which pattern retrieval is prevented by neurogenesis in the dentate gyrus.}
\label{schemaConc}
\end{figure}

\paragraph{Summary}
%The standard consolidation theory
%describes two interacting memory storage systems. The neocortex needs several days to consolidate its memory pattern through its different regions, but this memory can remain for years. The consolidation of the latter is ensured by the reactivations of the hippocampal engram which is fastly strengthened but also erased within a few weeks~\citep{squireRetrogradeAmnesiaMemory1995}.\\
Previous computational models have reproduced SCT processes and addressed a range of related fundamental questions~\citep{squireRetrogradeAmnesiaMemory1995,mcclellandWhyThereAre1995,
kaliOfflineReplayMaintains2004,meeterTracelinkModelConsolidation2005,
amaralSynapticReinforcementbasedModel2008,helferComputationalModelSystems2020a,
howardModelBidirectionalInteractions2022}.
However, to our knowledge, little attention has been paid to the underlying neurobiological processes and how they are responsible for the differences in learning rates and the erasure of memories between the neocortex and the hippocampus.
We proposed in this paper a computational model of SCT that embarks two main hypotheses from the neurobiological literature:
\begin{itemize}
\item  The slower consolidation in the neocortex could be explained by its larger structure, implying longer durations to connect remote areas of the same memory pattern.
\item The forgetting of hippocampal memories could be due to adult neurogenesis in the dentate gyrus and perturbation of memory retrieval.
\end{itemize}
%These two considerations, within a complex spatial structure of three interconnected brain areas (neocortex, dentate gyrus, and CA1 subfields) and following a temporal process composed of two kinds of important steps (hippocampal replay and retrieval cue) make the whole process challenging to model. This complexity, as well as the tissue-level scale of the theory, makes the neural field theory an appealing framework for SCT modelling. 
To model SCT, we adopted an original neural field approach, with original components compared to classic neural field models; such as interactions between three coupled neural fields or long-term synaptic plasticity. Our numerical simulations reproduced the main features of the standard consolidation theory, through a set of emerging processes that we summarize on figure~\ref{schemaConc}. 
%The neocortical pattern, initially dependent on the fastly constituted hippocampus for retrieval, becomes independent at the end of the consolidation, while the hippocampal pattern disappears.\\
%This computational model represents three coupled neural fields, with dynamic evolutions of neuronal thresholds and connections strengths, undergoing different neurobiological processes. 
%The structure evolves with time, undergoing progressive reorganization through repetitive steps (retrieval cue, hippocampal replay) and the effect of neurogenesis. 
After the initial encoding step, the connections between the elements of the neocortical memory pattern (bumps) are weaker than those in the hippocampus, because of the larger distance between them. Therefore, at this stage, pattern retrieval in the neocortex needs neocortex-hippocampus connections (see figure~\ref{schemaConc}~(a)): during neocortical pattern retrieval, neural activity is first transmitted from the stimulated neocortical element, here $B^C$, to its equivalent area in the hippocampus ($B^D$ and  $B^H$, steps 1-2 in fig.~\ref{schemaConc}~(a)). This subsequently activates the other parts of the hippocampal memory pattern as an intra-hippocampal process (step 3, $A^D$ and  $A^H$). These activations of $A^D$ and  $A^H$ eventually feedback to the neocortical neural field to retrieve the activation of $A^C$ (step 4). However as the consolidation process progresses, hippocampal replays as well as retrieval cue steps induce bumps of firing and thus learning weight consolidation in the three neural fields, even for connections located more remotely and in particular in the neocortex. At this stage, then, the connections between $A^C$ and $B^C$ have become large enough that the activation of $B^C$ during retrieval directly activates $A^C$, as a pure intra-cortical process (steps 1-2 of fig.~\ref{schemaConc}~(b)).

In parallel with the learning dynamics, we explored the effect of neurogenesis in the dentate gyrus on the stability of hippocampal memories. We modelled the insertion of highly excitable newborn neurons by a reduction of the threshold of a small fraction of dentate gyrus neurons in the neighborhood of the pattern neurons. The sustained firing of newborn neurons was found to prevent the reactivation of both dentate gyrus and CA memory patterns, due to local threshold adaptations in nearby pattern neurons. This inhibits hippocampal memory retrieval as soon as neurogenesis is significant, i.e. at long timescales, where the neocortical engram is the only one to be retrieved during the retrieval cue steps (see fig.~\ref{schemaConc}~(b)).

An important prediction of our model is that the structure of the trisynaptic pathway could play a crucial role in SCT. Indeed, we found out that a crucial element in our model is that the neural field where neurogenesis takes place (here the dentate gyrus) does not project connections to the neocortex, but to a third neural field (here the CA1-CA3 subfield region). In the presence of direct connections from the neurogenesis neural field to the neocortical one, the insertion of the newborn neurons erases the memory patterns not only in the hippocampus, but also in the neocortex. Hence, in our model, the connectivity structure of the trisynaptic pathway, with its three individual regions and a partially feed-forward connectivity (no connection from the dentate gyrus to the neocortex, in particular), allows neurogenesis to erase hippocampal memory patterns while sparing neocortical ones. 

We sketch in the supplementary material section~\ref{stability} an analytical framework to study the multi-bump pattern solutions in our model and their stability with regards to spike frequency adaptation, synaptic depression and long-term synaptic plasticity. This analysis is challenging because of the spatial and temporal dimensions of the bump pattern, that demands considering complicated spatio-temporal perturbations (contraction, expansion, shifts) and because of the presence of three interconnected neural fields. Because of this, we could not complete the analysis of all the possible perturbations. However, our preliminary analysis shows that the presence of learning weights (long-term synaptic plasticity) during the consolidation process stabilizes the multi-bump solution towards contraction perturbations whereas it has no effect on expansion perturbations. Future work will be needed to interpret these results in neurobiological terms, in particular regarding shift perturbations, but we believe that the possibility to produce such analytical evaluation even in the case of interacting complex neurobiological processes, is one of the main interest of neural fields.

\paragraph{Future directions}
Despite its complexity, our computational model is still highly simplified in regards to the neurobiology and could be improved in several ways.
The modelling of the learning weights (eq.~\eqref{eq:wtot} to~\eqref{eq:etas}) could be refined to achieve expressions with a stronger biophysical support, such as the adaptation of STDP for neural fields~\citep{robinsonNeuralFieldTheory2011}, while maintaining the continuum in time and space. In particular the phenomenological continuity function $\eta_s^{\alpha\beta}(x,t)$ (eq.~\eqref{eq:etas}) could be re-evaluated, and replaced by another mechanism providing continuity of the bump solutions, an important feature in neural field models. Furthermore, to approach biological reality, it is important to question the pertinence of the use of neural fields for hippocampus networks since neural fields have been developed to model neocortical networks~\citep{coombesWavesBumpsPatterns2005}.
Here, we used an homogeneous external input to induce hippocampal replays, thus emulating slow-wave sleep. This could be replaced by a more realistic spontaneous reactivation mechanism~\citep{klinzingMechanismsSystemsMemory2019}.
%It would be compelling to adapt the model for stochastic neural fields or for spiking neural networks. \\

Moreover, the timescales used in our model are questionable. We take for reference the time constant for the neural activity $u^\alpha(x,t)$ which has been fixed to $1$ in our model, but which is estimated to be around $10$ ms experimentally~\citep{abbottSynapticDepressionCortical1997}. 
In our model the synaptic resources recovery rate $\alpha_q$ in synaptic depression eq.\eqref{eq:qdiff}, should lie between $20$ and $80$ (since experimentally estimated to be between $200$  and $800$ ms~\citep{abbottSynapticDepressionCortical1997}). However, the numerical value used in our simulations is $\alpha_q=800$. Likewise, the spike frequency timescale $\tau_\kappa$ should be between $4$ and $12$ (experimentally estimated to be between $40$ and $120$ ms~\citep{madisonControlRepetitiveDischarge1984}). But in our numerical simulations we used $\tau_\kappa=0.8$. These discrepancies between the experimental values and those used for our parameters can be explained by the complexity of the process to model. Indeed multiple timescales  are present in the model, including those mentioned above, the activity-dependent synaptic resources depletion rate $m_q^{\alpha}(x,t)$ (equations\eqref{eq:qdiff}\eqref{eq:mq}), the timescale for the threshold related to neurogenesis $\tau_{\kappa_n}$ (eq.\eqref{eq:kndiff}), the distance-dependent learning rate $d(x,y)$ (equations\eqref{eq:LweightsL}\eqref{eq:dLR}), the forgetting rate $c_0$ (eq.\eqref{eq:LweightsS}), the duration of the stimulations $I_\text{ext}^\alpha(x,t)$ (eq.\eqref{eq:IextC} for $I_\text{ext}^C(x,t)$). The values of these time-scale parameters must be reevaluated in future works, to better approach biological reality.

Furthermore, additional phenomenological mechanisms could be accounted for. In the equation for the threshold related to neurogenesis ${\kappa_n}^\alpha(x,t)$,  eq.\eqref{eq:kndiff}, we used a Heaviside function for simplicity. But the effects of a smoother firing rate function would need to be studied in the future. In addition, the phenomenological introduction of an activity dependent synaptic resources depletion rate $m_q^\alpha(x,t) = \beta_q u^\alpha(x,t)$ in equations\eqref{eq:qdiff}\eqref{eq:mq} was found convenient in our model because it allows to get a roughly constant duration for the bump patterns even with changing synaptic weights or facing neurogenesis. However further work must be carried out to study whether this mechanism has a physiological meaning or should be removed from the model.

This model succinctly studied the neurogenesis mechanism leading to the erasure of memories, and would need to be further developed to be fully understood as it has been in previous modelling works~\citep{deisserothExcitationNeurogenesisCouplingAdult2004,meltzerRoleCircuitHomeostasis2005,weiszNeurogenesisInterferesRetrieval2012, beckerComputationalPrincipleHippocampal2005}. In particular, its integration in a neural field model in coordination with the different homeostasis processes could be investigated in more detail. Another interesting hypothesis for forgetting is evoked by experiments of silencing of engrams~\citep{tonegawaRoleEngramCells2018,josselynMemoryEngramsRecalling2020}.

Finally our model aimed at providing a proof of principle and a selection for the main ingredients to include in the model. It would be interesting to test whether similar results can be obtained with a more realistic neural network, such as a stochastic spiking neural network. Such a model would allow more complex pattern configurations and more accurate dynamics rules for the learning weights. More detailed attention to anatomical properties of the circuits would also be of great interest~\citep{pykaPatternAssociationConsolidation2014} and more realistic synaptic plasticity equations could be implemented~\citep{tomeCoordinatedHippocampalthalamiccorticalCommunication2022}. Thus, it would be compelling to derive a neural field model from such a microscopic model, and to compare its components with those developed in this paper.\\
%\citep{guanHowDoesSparse2016}\\

%\textbf{NF ideas to include on the other paragraphs}
%Effect of learning weights on other dynamics than bumps, such as traveling waves or turing patterns.
%
%inhomogeneous media<->LW?
%
%Heterogeneous jirsa, 
%
%multibumps
%
%plasticity stdp adapted
%
%make the link with topo maps
%
%Link with working memory PFC which has already been studied in NF
%
%stochastic
%
%inverse modelling

%
%
%\textbf{*Systems consolidation theories}

Here, we focused on the widespread standard consolidation theory. However, there exist other models and questions around systems memory consolidation concepts. For instance, the multiple trace theory suggests that some of the hippocampus patterns are conserved in the long term. This theory follows observations of hippocampal damages that produced temporally-graded retrograde amnesia only for semantic memories, but not for episodic ones~\citep{nadelMemoryConsolidationRetrograde1997}.
The trace transformation theory posits a selection process that depends on the circumstances at retrieval, and permits the persistence of both the neocortical or the hippocampal memory~\citep{winocurMemoryFormationLongterm2010}. On the other hand, the more recent concept of active systems consolidation studies in greater detail the influence of sleep on consolidation~\citep{klinzingMechanismsSystemsMemory2019}. Another interesting related phenomenon which has already been modelled in some connectionist models~\citep{amaralSynapticReinforcementbasedModel2008,helferComputationalModelSystems2020a}, is systems memory reconsolidation. In this approach, it is the replay of an already consolidated memory, which can involve again the hippocampus~\citep{naderMemoryReconsolidationUpdate2010}.\\

In addition to these theories, some models suggest that the prefrontal cortex might play a key role in memory organization~\citep{larochePlasticityHippocampalPrefrontal2000}.
On the one hand, the prefrontal cortex is involved in the processing and integration of ancient neocortical memories and seems to inhibit hippocampal activity when new information is too similar to an already stored neocortical pattern~\citep{franklandOrganizationRecentRemote2005,prestonInterplayHippocampusPrefrontal2013}.
This has been included in a recent connectionist model~\citep{hwuNeuralModelSchemas2020}. On the other hand, the prefrontal cortex has also been identified to be a location for working memory~\citep{joaquinCognitiveFunctionsPrefrontal2010}. A link between working and long-term memory has been introduced in a modelling study, with the hippocampus modelled as intermediate-term memory~\citep{fiebigMemoryConsolidationSeconds2014}.
Tonegawa et al, also studied the role of the prefrontal cortex and the basolateral amygdala in memory reorganization~\citep{tonegawaRoleEngramCells2018}. The implementation of all these proposals into a coupled neural field model similar to the one used here could open perspectives in the domain.

Finally, the preliminary stability analysis results shown in supplementary materials~\ref{appendices} have not been adressed numerically through the different perturbations. It remains an important issue to adress in the future, in particular because our mathematical analysis only provides conditions for instability, not for stability. Furthermore, for the analysis of the synaptic depression, we used $m_q^\alpha(x,t) = \beta_q$, but it will be important to study the case $m_q^\alpha(x,t) = \beta_q u^\alpha(x,t)$, as used in our numerical simulations.

 \bibliographystyle{elsarticle-harv} 
 \bibliography{biblioJTB}





% \end{thebibliography}

\newpage
\appendix

\setcounter{figure}{0}

\section{Supplementary text}
\label{appendices}
\subsection{Effect of learning weights on bump solutions and stability analyses}
\label{stability}
This subsection focuses on the effect of the learning weights on the existence and the stability of stationary bumps solutions. We assume that the width of the bump is conserved across the neural fields, i.e. $a^C=a^D=a^H=a$, which is verified numerically. To carry out an analytical treatment of existence and stability of the bump patterns, we consider the case where all learning weights are at their equilibrium value, after the consolidation process, with the decay term $F^{\alpha\beta}(x,y,t)$ neglected, since $c_0<< \text{min}(d(x,y))$. In fact this leads to a dimension reduction of the system, since bump profiles will all be identical. Which is not the case for transient learning weights due to the differences of timescales within and between neural fields. Furthermore, the threshold $\kappa_n^\alpha(x,t)$ is not considered here, due to its large time scale. It could be used as a slow-varying parameter in continuation bifurcation analysis.
The effects of spike frequency adaptation and synaptic depression are studied separately below. 

Throughout this analysis we will use the results of Kilpatrick and Bressloff~\citep{kilpatrickStabilityBumpsPiecewise2010, kilpatrickEffectsSynapticDepression2010}, that are restricted to a single field, but we adapt them to our model composed of three connected neural fields, with learning
weights.
%This analysis we will mostly benefit from the results of Kilpatrick and Bressloff~\citep{kilpatrickStabilityBumpsPiecewise2010, kilpatrickEffectsSynapticDepression2010}. Their studies are restricted to a single neural field, but we adapt them to our model composed of three inter-connected neural fields, with learning weights.

\subsubsection{Spike frequency adaptation}
\label{spkanalysis}
\paragraph{Existence of stationary bumps solution}
As is usual in the field of neural fields, we replace all the continuous firing functions $f$ and $f_a$ by discrete Heaviside functions ($\beta_f\xrightarrow{}\infty$, $\beta_{f_a}\xrightarrow{}\infty$).

In this case a stationary bump solution of our model $\bigl(u_0^\alpha(x),\kappa_0^\alpha(x),s_0^{\alpha\beta}(x,y)\bigr)$ satisfies:

\begin{equation}
u_0^\alpha(x) =
\begin{cases}
\sum\limits_{\beta\in E}G^{\alpha\beta}J^{\alpha\beta}(x,a) & \text{ if } x\notin R[u_0^\alpha]\\
\sum\limits_{\beta\in E}G^{\alpha\beta}[J^{\alpha\beta}(x,a)\xi_s - (\xi_s-1)\kappa_\text{in}] & \text{ if } x\in R[u_0^\alpha]
\end{cases},
\label{eq:u0spk}
\end{equation}

\begin{equation}
\kappa^\alpha_0(x) =
\begin{cases}
\kappa_\text{in} & \text{ if } x\notin R[\kappa_0^\alpha]\\
\kappa_\text{in}+\eta_\kappa & \text{ if } x\in R[\kappa_0^\alpha]
\end{cases},
\label{eq:k0spk}
\end{equation}
and
\begin{equation}
s_0^{\alpha\beta}(x,y) =
\begin{cases}
0 & \text{ if } x\notin R[u_0^\alpha] \text{ or } y\notin R[u_0^\beta]\\
1 & \text{ if } x\in R[u_0^\alpha]\text{, }y\in R[u_0^\beta]
\end{cases}
\label{eq:w0spk}
\end{equation}

with

\begin{equation}
J^{\alpha\beta}(x,a) = \sum_{z\in Z^\beta}\biggl[W(x-\Delta^{\alpha\beta}(x)-(z-a))-W(x-\Delta^{\alpha\beta}(x)-(z+a))\biggr].
\end{equation}

$R[u_0^\alpha]$ is the excited region for $u_0^\alpha$, which corresponds also to the location where learning weights have a non-zero value, defined as

\begin{equation}
R[u_0^\alpha] = \bigcup_{z\in Z^\alpha}[z-a,z+a]
\label{eq:Ru0}
\end{equation}
The bump boundaries are defined by the threshold conditions:

\begin{equation}
u_0^\alpha(z \pm a) = \kappa_\text{in}+\eta_\kappa = 2\times 2ae^{-2a/s}
\label{eq:u0BC}
\end{equation}

$R[\kappa_0^\alpha]$, the excited region for $\kappa_0^\alpha$ is different

\begin{equation}
R[\kappa_0^\alpha] = \bigcup_{z\in Z^\alpha}[z-b,z+b]
\label{eq:Rk0}
\end{equation}
with $b$ implied in the relation

\begin{equation}
u_0^\alpha(z \pm b) = \kappa_\text{in} =  2ae^{-2a/s}
\end{equation}
Please note that $b>a$. The factor $\xi_s$ is defined as:

\begin{equation}
\xi_s = 1+\gamma(1+{cr})4a
\end{equation}
The second term in the right-hand side of this equation represents the sum over the neural fields and the bumps of the integral of the learning weights. The term $\xi_s$ accounts for the learning weights in the neural fields model. In the absence of learning, $\xi_s=1$, otherwise $\xi_s>1$. We will study the effect of varying $\xi_s$ on the existence of bumps and on their stability.
We introduce the total weights defined as the sum of the synaptic plasticity term with learning weight at equilibrium and the permanent weights.

\begin{equation}
w_\text{tot}^{\alpha\beta}(x,y)=w(x-\Delta^{\alpha\beta}(x)-y)+\gamma^{\alpha\beta}\eta_{s}^{\alpha\beta}(x)s_0^{\alpha\beta}(x,y)
\end{equation}
where 

\begin{equation}
\eta_{s}^{\alpha\beta}(x) = 
\biggl[J^{\alpha\beta}(x,a) - \kappa_\text{in}\biggr] _+
\end{equation}

\begin{figure}
\centering
\includegraphics[width=6cm]{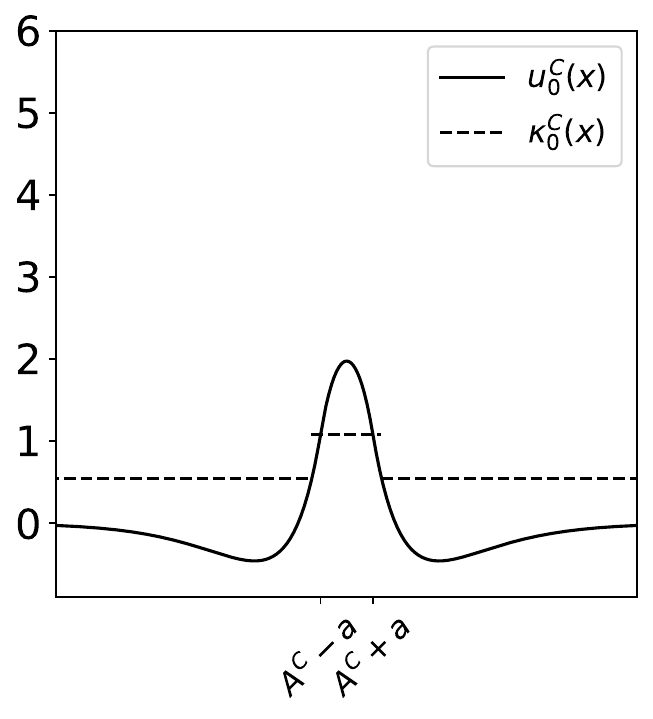}
\includegraphics[width=6cm]{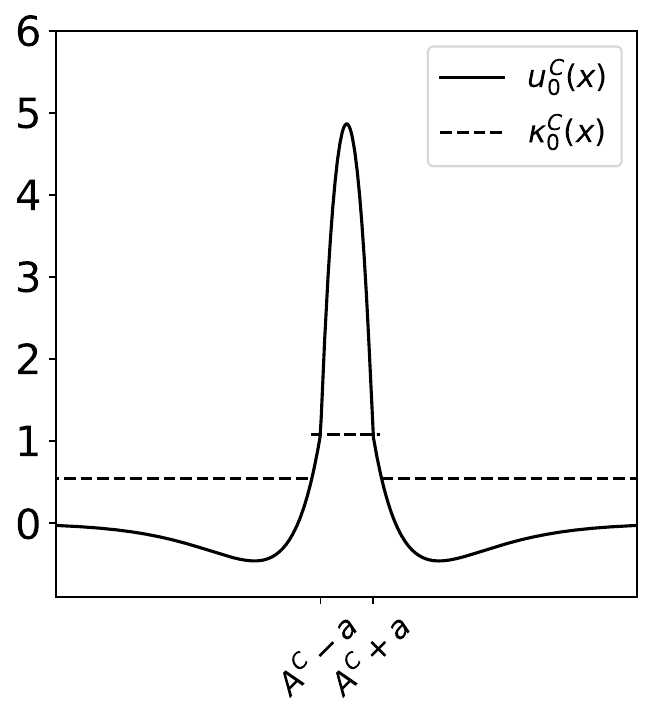}
\caption[Shape of a bump in the C field]{Shape of a bump in the C neural field in two extreme cases: (\textit{left}) in the absence of learning, i.e. for $s_0^{\alpha\beta}(x,y)=0$ along the whole field and (\textit{right}) with synaptic modifications at their maximum values, i.e. for $s_0^{\alpha\beta}(x,y)$ set to 1 in the pattern areas.}
\label{bumps}
\end{figure}

Two bump profiles in the C neural field with all the $s_0^{\alpha\beta}(x,y)$ set to 0 or to 1, are presented in figures~\ref{bumps}~(Left) and~(Right), respectively. The figure confirms that the continuity of the solution is preserved even with learning weights. The bump boundaries $a$ are not modified by the introduction of synaptic plasticity, but inside the $R[u_0^C]$ zone the bump amplitude is larger with synaptic plasticity.

\paragraph{Stability of the bumps}
Following the computations of Kilpatrick and Bressloff (2010)~\citep{kilpatrickStabilityBumpsPiecewise2010}, we develop eq.~\eqref{eq:udiff} with a spatio-temporal perturbation, $u^\alpha(x,t) = u_0^\alpha(x) + \epsilon \phi_u^\alpha(x,t)$, where $\phi_u^\alpha(x,t)$ are smooth perturations and $\epsilon<<1$. Since we truncate the results at first order, no learning weights nor threshold perturbations will appear. The linear stability analysis we derive here thus does not reflect the underlying translation invariance of the system. \\
We then assume time-space separability of the perturbation: $\phi_u^\alpha(x,t) = e^{\lambda t}\psi_u^\alpha(x)$. The calculations are detailed in Ref.~\citep{kilpatrickStabilityBumpsPiecewise2010}, the modifications of our analysis compared to this study is mainly that the value above which the threshold starts increasing and the baseline threshold are both equal to $\kappa_\text{in}$, which simplifies the analysis. Moreover, in our model,
the interconnections between the three neural fields and the learning weights add a $2$ factor multiplying the permanent weights, and induce a modification of the spatial derivative of $u_0^\alpha(x)$. In fact, the factor $\pi_u$ takes into account different left and right derivatives (see below).\\
The general perturbation equation can be written as:
\begin{equation}
\begin{split}
(\lambda+1)\psi_u^\alpha(x) = \sum_{\beta\in E}G^{\alpha\beta}\sum_{z\in Z^\beta}\Biggl[& w_\text{tot}^{\alpha\beta}(x,z-a)\frac{\psi_u^\beta(z-a)}{|u_0'^\beta(z- a)|}+ w_\text{tot}^{\alpha\beta}(x,z+a)\frac{\psi_u^\beta(z+a)}{|u_0'^\beta(z+ a)|} \Biggr]
\end{split}
\end{equation}
with
\begin{equation}
\frac{1}{|u_0'^\alpha(z+\sigma a)|}=
\begin{cases}
\pi_u \text{ if } \psi_u^\alpha(z+\sigma a)>0\\
\frac{\pi_u}{\xi_s} \text{ if } \psi_u^\alpha(z+\sigma a)<0
\end{cases}
\label{eq:guspk}
\end{equation}
and

\begin{equation}
\pi_u = \frac{1}{2(w(0)-w(2a))}
\end{equation}

The essential spectrum is located at $\lambda = -1$. The discrete spectrum is obtained by setting $x = z\pm a$, with $z=A^\alpha$ or $B^\alpha$. At these boundaries $\eta_s^\alpha(x)=0$, erasing the learning weights terms. Furthermore the distance between two bumps, even in the nearest case in $D,H$, is large enough so that the values of the non-learned, permanent weights between them can be  neglected. Moreover, all learning weights are equal. Therefore, we can assume that $u_0^\alpha(x)$ and $\kappa_0^\alpha(x)$ are equal at their bump boundaries for all the neural field. We also consider that $\psi_u^\alpha(x)$ and $\psi_\kappa^\alpha(x)$ exhibit identical values at each boundary. This simplification allows the calculations below. Moreover, within the linear regime, infinitesimal changes in $u^\alpha$ will only perturb the threshold in a neighborhood of $x = z\pm b$, so that $\phi_\kappa^\alpha(z\pm a,t) = 0$.\\
There are four classes of solutions which determine the discrete spectrum: expansion, contraction, leftward shift and rightward shift of the stationary bump solution.
\paragraph{Expansion case:}
$\forall\beta\in E, \forall z\in Z^\beta, \forall \sigma\in\{-1,1\}$, $\psi_u^\beta(z+\sigma a)>0$

\begin{equation}
\biggl[2\pi_u[w(0)-w(2a)]-(\lambda+1)\biggr]\biggl[2\pi_u[w(0)+w(2a)]-(\lambda+1)\biggr] = 0
\end{equation}

\begin{equation}
\lambda_\pm = \frac{w(0)\pm w(2a)}{w(0)- w(2a)}-1
\end{equation}

These eigenvalues are independent of $\xi_s$. With our parameter values, we find that $\lambda_-=0$ and $\lambda_+>0$, uncovering a degenerate case for which we cannot conclude on the stability.

\begin{figure}[!h]
\centering
\includegraphics[width=8cm]{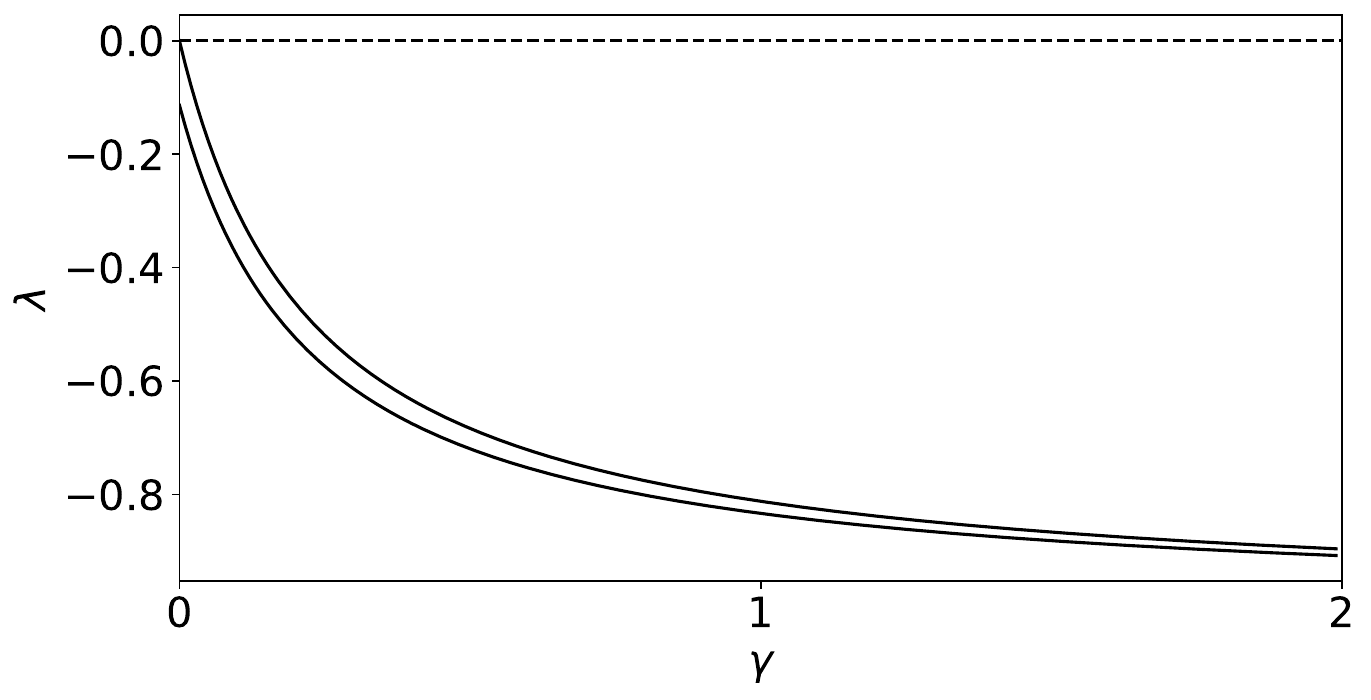}
\caption[Learning weights stabilize the bump in case of a contraction perturbation]{Learning weights stabilize the bump in case of a contraction perturbation. Eigenvalues $\lambda$ decrease as $\gamma$ increases (which increases $\xi_s$, as $\xi_s = 1+\gamma(1+{cr})4a$).}
\label{lSpk}
\end{figure}

\paragraph{Contraction case:}
$\forall\beta\in E, \forall z\in Z^\beta, \forall \sigma\in\{-1,1\}$, $\psi_u^\beta(z+\sigma a)<0$

\begin{equation}
\lambda_\pm = \frac{1}{\xi_s}\frac{w(0)\pm w(2a)}{w(0)- w(2a)}-1
\end{equation}
For $\xi_s>1$, $\lambda_\pm<0$.
Figure~\ref{lSpk} presents the evolution of the contraction eigenvalues with $\gamma$ ($\xi_s = 1+\gamma(1+{cr})4a$). With our parameter values the bump is stable to contraction perturbation, with increasing stability when $\gamma$ increases. 

\paragraph{Leftward shift case:}
This case corresponds to a class of solutions where $\forall\beta\in E, \forall z\in Z^\beta,$ $\psi_u^\beta(z+a)<0$ and $\psi_u^\beta(z-a)>0$.

\begin{equation}
\biggl[2\pi_u w(0)-(\lambda+1)\biggr]\biggl[2\pi_u /\xi_s w(0)-(\lambda+1)\biggr]-1/\xi_s \biggl[2\pi_u w(2a)\biggr]^2 = 0
\end{equation}

\begin{equation}
\lambda_\pm = 2\pi_u\frac{(1+1/\xi_s)w(0)\pm \sqrt{(1+1/\xi_s)^2w(0)^2-4[w(0)^2-1/\xi_s w(2a)^2]}}{2} -1
\end{equation}

\begin{figure}[!h]
\centering
\includegraphics[width=8cm]{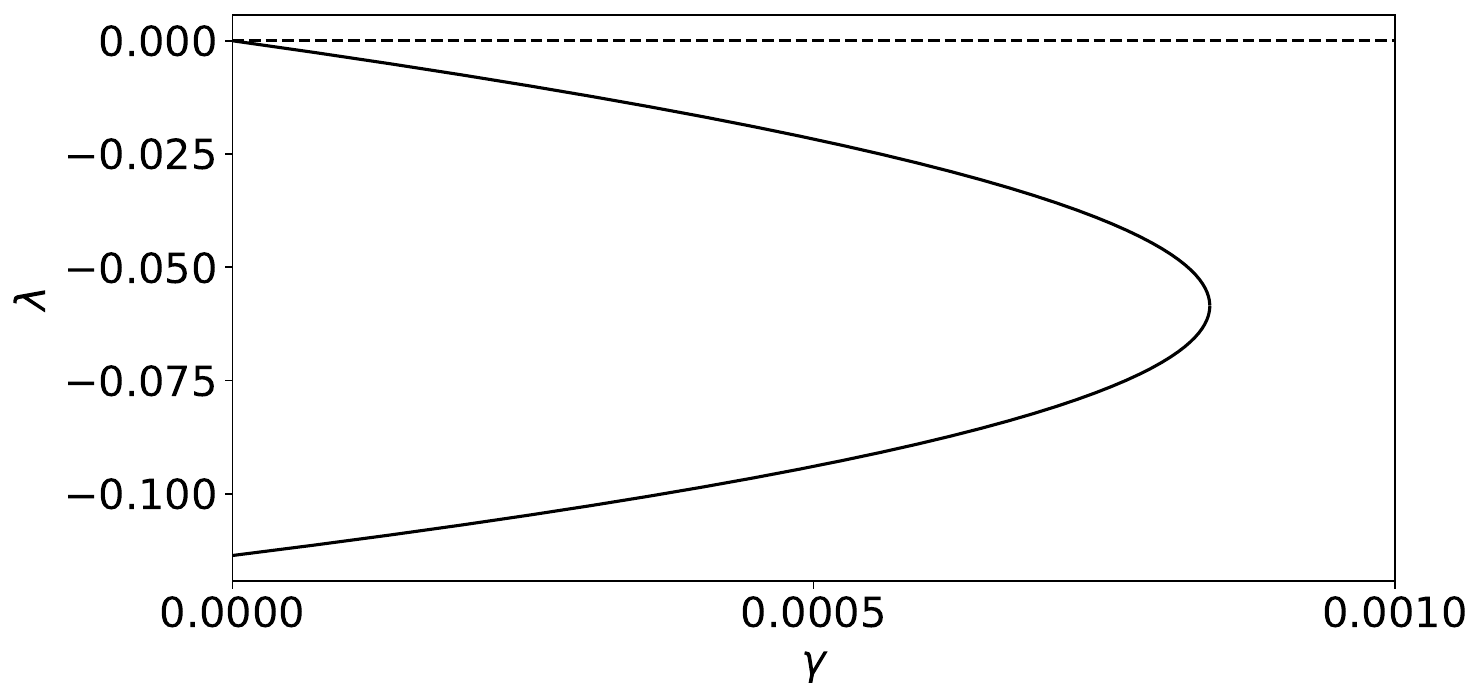}
\caption[It is not possible to conclude on the influence of learning weights in case of a shift perturbation]{It is not possible to conclude on the influence of learning weights in case of a shift perturbation. These eigenvalues are real for small $\gamma$ values, where $\lambda_\pm\leq 0$. $\lambda_+$ decreases and $\lambda_-$ increases as $\gamma$ increases (which increases $\xi_s$, as $\xi_s = 1+\gamma(1+{cr})4a$).}
\label{SpkgammaShift}
\end{figure}

These eigenvalues are real for small $\xi_s$ values (complex eigenvalues violate the initial assumption that $\lambda$ are real), where $\lambda_\pm< 0$ for $\xi_s>1$. When $\xi_s$ increases, $\lambda_+$ decreases and $\lambda_-$ increases, see figure~\ref{SpkgammaShift}. Therefore, we cannot conclude on the effect of learning weights on shift perturbations.

\paragraph{Rightward shift case:}
This case corresponds to a class of solutions where $\forall\beta\in E, \forall z\in Z^\beta,$ $\psi_u^\beta(z+a)>0$ and $\psi_u^\beta(z-a)<0$.\\
The spectrum associated with rightward shifts is identical to that of leftward
shifts due to the reflection symmetry of the system.\\

To conclude, the introduction of synaptic plasticity / learning weights stabilizes the system through contraction, but has no effects on stability in case of an expansion perturbation. We cannot conclude on their influence on the shift behaviour since it increases one eigenvalue but diminishes the other one. This method can only provide sufficient conditions for instability but not for stability of a bump. So there could be unstable modes not detected by this analysis, although numerical simulations suggest that this is not the case for the parameters of the model.

\subsubsection{Synaptic depression}
We now study the effect of the learning weights on the existence of bumps and stability in presence of synaptic depression. With the parameters values used in our model, there is no stable / stationary bump solution when synaptic depression is added. Synaptic depression therefore shuts down the neural bump activities. We still develop the analysis, which could be used in the future to study the effect of the learning weights on an alternative system, but we do not compute values of eigenvalues and thus, we do not conclude here on stability behaviours. We consider the original equation with
$m_q^\alpha(x,t) = \beta_q$.

\paragraph{Existence of stationary bumps solution}
After replacement of the firing functions by heaviside functions,
a stationary bump solution $\bigl(u_0^\alpha(x),q_0^\alpha(x),s_0^{\alpha\beta}(x,y)\bigr)$ is found to satisfy:

\begin{equation}
u_0^\alpha(x) =
\begin{cases}
\frac{1}{1+\alpha_q\beta_q}\sum\limits_{\beta\in E}G^{\alpha\beta}J^{\alpha\beta}(x,a) & \text{ if } x\notin R[u_0^\alpha]\\
\frac{1}{1+\alpha_q\beta_q}\sum\limits_{\beta\in E}G^{\alpha\beta}[
J^{\alpha\beta}(x,a)\xi_s - (\xi_s-1)\kappa_\text{in}] & \text{ if } x\in R[u_0^\alpha]
\end{cases}
\label{eq:u0syn}
\end{equation}

The excited region and bumps boundaries are defined by the same conditions described in equations\eqref{eq:Ru0} and\eqref{eq:u0BC}.
%\begin{equation}
%u_0^\alpha(z \pm a) = \kappa_{in}+\eta_\kappa
%\end{equation}

\begin{equation}
q_0^\alpha(x) = \begin{cases}
1 & \text{ if } x\notin R[u_0^\alpha] \\
\frac{1}{1+\alpha_q\beta_q} & \text{ if } x\in R[u_0^\alpha]
\end{cases}
\label{eq:q0syn}
\end{equation}

The equilibrium solution of the learning weight factor $s_0^{\alpha\beta}(x,y)$ was defined in\eqref{eq:w0spk}. We remind that for our parameter values, this bump solution does not exist.

\paragraph{Stability of the bumps} Following Kilpatrick and Bressloff ~\citep{kilpatrickStabilityBumpsPiecewise2010}, we develop eq.\eqref{eq:udiff} and\eqref{eq:kdiff} with $u^\alpha(x,t) = u_0^\alpha(x) + \epsilon \phi_u^\alpha(x,t)$ and $q^\alpha(x,t) = q_0^\alpha(x) + \epsilon \phi_q^\alpha(x,t)$, where $\psi_u^\alpha(x,t)$, $\phi_q^\alpha(x,t)$ are smooth perturations and $\epsilon<<1$. 
With a method similar to the one developed in the previous subsubsection~\ref{spkanalysis}, we find the general equation:

\begin{equation}
\begin{split}
&(\lambda+1)\psi_u^\alpha(x) = \\&
\sum_{\beta\in E}G^{\alpha\beta} \sum_{z\in Z^\beta}\sum_{\sigma\in\{-1,1\}}w_\text{tot}^{\alpha\beta}(x,z+\sigma a)\biggl[1-\frac{\beta_q\Theta(\psi_u^\beta(z+\sigma a))}{(\lambda+1/\alpha_q+\beta_q)} \biggr]G(\psi_u^\beta(z+\sigma a))\frac{\psi_u^\beta(z+\sigma a)}{|u_0'^\beta(z+\sigma a)|}
\end{split}
\label{eq:closedE}
\end{equation}

$\frac{1}{|u_0'^\alpha(z+\sigma a)|}$ is defined as in eq.\eqref{eq:guspk}, but with $\pi_u$ defined here as:

\begin{equation}
\pi_u = \frac{1+\alpha_q\beta_q}{2(w(0)-w(2a))}
\end{equation}

The function $G(X)$ is written as:

\begin{equation}
G(X) = \Biggl\{
\begin{array}{ll}
1 \text{ if } X>0\\
(1+\alpha_q\beta_q)^{-1} \text{ if } X<0
\end{array}
\end{equation}

Here again, there are four classes
of solutions which determine the discrete spectrum: expansion, contraction, leftward
shift and rightward shift of the
stationary bump solution.
\paragraph{Expansion case:}
This case corresponds to a class of solutions where $\forall\beta\in E, \forall z\in Z^\beta, \forall \sigma\in\{-1,1\}$, $\psi_u^\beta(z+\sigma a)>0$.

With the simplified equation, since all $\psi_u^\beta(z+\sigma a)$ have the same sign, all $\psi_u^\beta(z+\sigma a)$ are equal and all $\psi_u^\beta(z+\sigma a)>0$.

Finally the eigenvalue $\lambda$ satisfies the equation

\begin{equation}
(\lambda+\alpha_q^{-1}+\beta_q)(\lambda+1) = (\lambda+\alpha_q^{-1}) (1+\alpha_q\beta_q)\Omega
\end{equation}
With

\begin{equation}
\Omega = \frac{w(0)+w(2a)}{w(0)-w(2a)} 
\end{equation}
The solutions of this equation are:

\begin{equation}
\lambda_\pm = \frac{\Omega(1+\alpha_q\beta_q)-(1+\alpha_q^{-1}+\beta_q)\pm\sqrt{(\Omega(1+\alpha_q\beta_q)-(1+\alpha_q^{-1}+\beta_q))^2+4(\Omega-1)(\alpha_q^{-1}+\beta_q)}}{2}
\end{equation}

These eigenvalues are independent of $\xi_s$, thus stability does not depend on $\gamma$. 
%We find with our parameters values $\lambda_\pm>0$.
\paragraph{Contraction case:}
This case corresponds to a class of solutions where $\forall\beta\in E, \forall z\in Z^\beta, \forall \sigma\in\{-1,1\}$, $\psi_u^\beta(z+\sigma a)<0$.

With the simplified equation, since all $\psi_u^\beta(z+\sigma a)$ have the same sign, all $\psi_u^\beta(z+\sigma a)$ are equal and all $\psi_u^\beta(z+\sigma a)<0$. The solution is

\begin{equation}
\lambda =\Omega/\xi_s - 1
\end{equation}
Assuming $\Omega>0$ and $\xi_s\geq 1$, the stability is increased by learning weights when $\gamma$ increases (since $\xi_s = 1+\gamma(1+{cr})4a$). 

\paragraph{Leftward shift case:}
This case corresponds to a class of solutions where $\forall\beta\in E, \forall z\in Z^\beta,$ $\psi_u^\beta(z+a)<0$ and $\psi_u^\beta(z-a)>0$. With the simplified equation, since all $\psi_u^\beta(z+a)$ (respectively $\psi_u^\beta(z-a)$)  have the same sign, all $\psi_u^\beta(z+ a)$ ($\psi_u^\beta(z-a)$) are equal.

The resulting equation is
\begin{equation}
\begin{split}
&\biggl[\Gamma_{\beta_q}(\lambda) -\pi_u(\lambda+\alpha_q^{-1})2w(0)\biggr]\biggl[\Gamma_{\beta_q}(\lambda) -\pi_u/\xi_s(\lambda+\alpha_q^{-1})2w(0)\biggr] - 1/\xi_s\biggl[\pi_u(\lambda+\alpha_q^{-1})2w(2a)\biggr]^2\\& + \frac{\lambda\alpha_q\beta_q/\xi_s\gamma_{u}}{1+\alpha_q\beta_q}\biggl[[\Gamma_{\beta_q}(\lambda) -\pi_u(\lambda+\alpha_q^{-1})2w(0)]2w(0) - [\pi_u(\lambda+\alpha_q^{-1})2w(2a)]2w(2a)\biggr] = 0,
\end{split}
\label{psiMatrix}
\end{equation}
with
\begin{equation}
\Gamma_{\beta_q}(\lambda) = (\lambda+\alpha_q^{-1}+\beta_q)(\lambda+1)
\end{equation}
 
The roots of this equation can be searched for numerically, and must satisfy the conditions $\psi_u^\beta(z+a)<0$ and $\psi_u^\beta(z-a)>0$. But, since we do not compute them for any set of parameters, we cannot conclude on the effect of learning weights on stability in this case. 

\paragraph{Rightward shift case:}
This case corresponds to a class of solutions where $\forall\beta\in E, \forall z\in Z^\beta,$ $\psi_u^\beta(z+a)>0$ and $\psi_u^\beta(z-a)<0$. Due to the symmetry of the system, the spectrum
associated with rightward shifts is identical to that of leftward
shifts.

As a conclusion, even if we do not compute the eigenvalues with a given set of parameters, it was still possible to estimate the stability behaviours, provided that equilibrium bump solutions exist.
As in the previous subsubsection~\ref{spkanalysis}, we found that learning weights stabilize the system through contraction, but have no effects on stability in case of an expansion perturbation. However, we cannot conclude on their influence on the shift behaviour since we did not compute eigenvalues values.

All in all, this analysis provided an analytical framework to study the three neural fields model with spike frequency adaptation, synaptic depression and learning weights.

\newpage
\section{Supplementary figures}
\setcounter{figure}{0}
\begin{figure}[!ht]
\centering
\includegraphics[height=4.8cm]{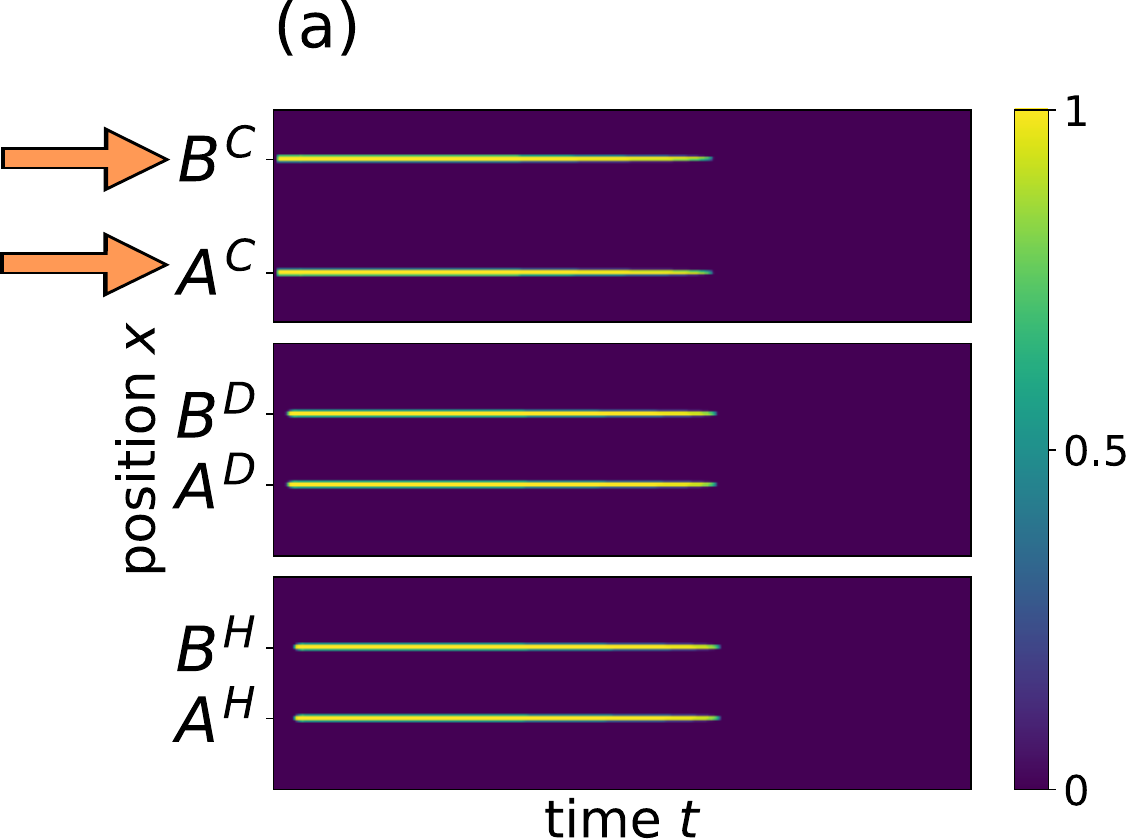}
\includegraphics[height=5cm]{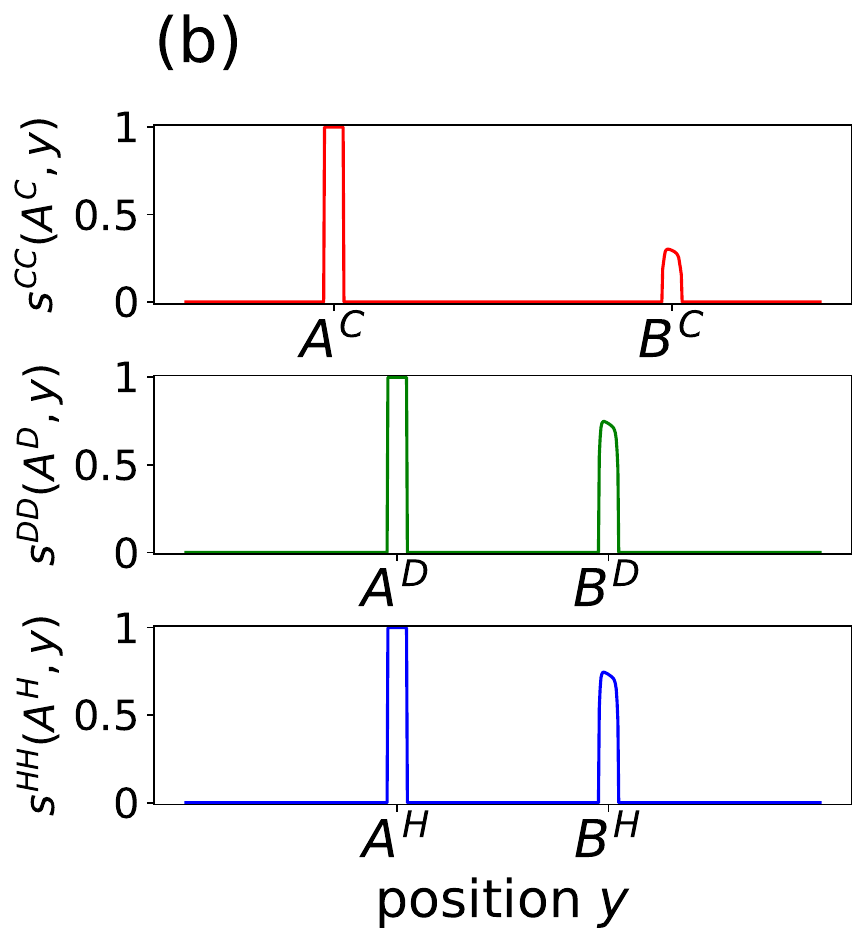}
\includegraphics[height=4cm]{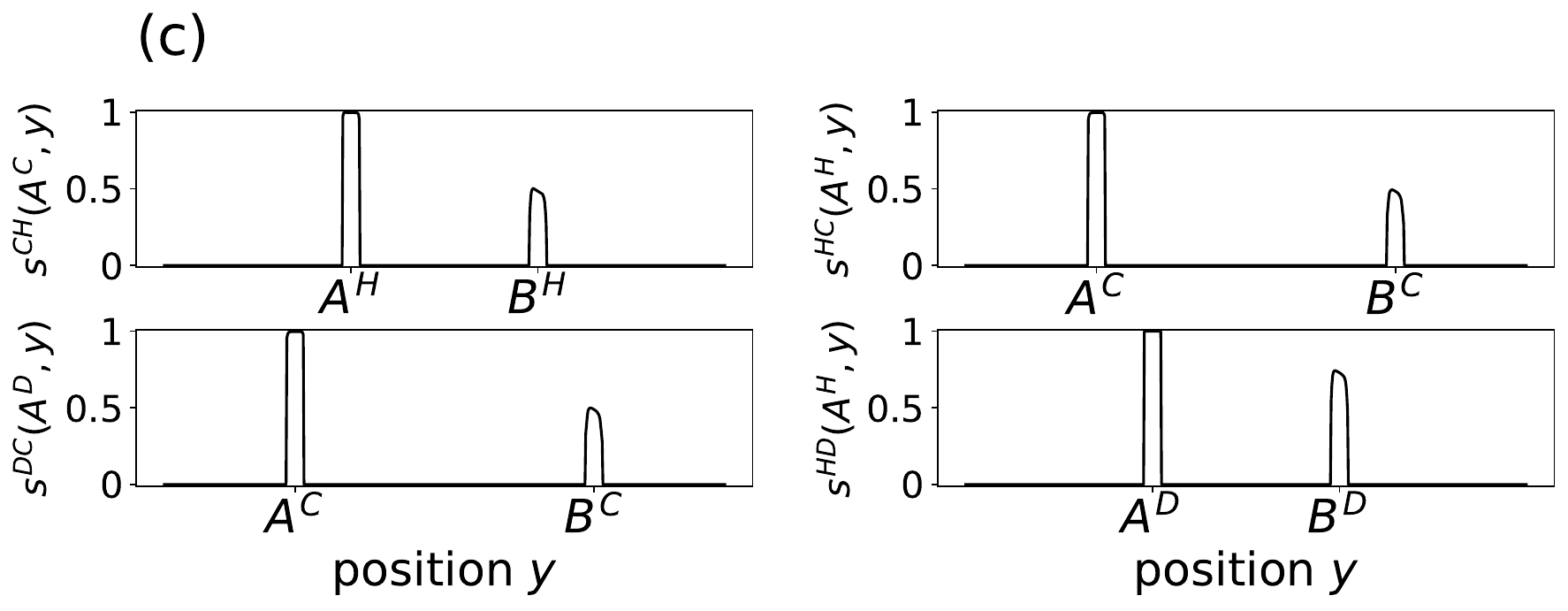}
\caption[The encoding step shows faster learning in hippocampal fields than in neocortical ones]{The encoding phase shows faster learning in hippocampal fields than in neocortical ones.  \textbf{(a)} Firing rates in the three fields, neocortex (C), dentate gyrus (D) and CA regions (H).  The initial stimuli are two bumps around positions $A^C$ and $B^C$ in the neocortical field. They activate C neurons which in turn activate D neurons. Then since C and D neurons fire, H neurons can be activated. In the end, neurons shutdown, due to the depletion of synaptic resources. \textbf{(b)} Intra and \textbf{(c)} inter-field learning weights at the end of the encoding phase. Learning weights within the $A^\alpha$ part of each pattern are at their maximum. Indeed, neurons inside this location are very close to each other, so their weights grow fast. On the opposite the weights between the $A^\alpha$ and $B^\alpha$ parts are smaller, since the distance is more important. Especially these cross weights are still smaller in C than in D and H fields, since the distance between $A^C$ and $B^C$ is larger than those between $A^D$ and $B^D$ (or $A^H$ and $B^H$). Thus in our model, distance is the main reason why the neocortex is a slow learner and the hippocampus a fast learner. }
\label{encoding}
\end{figure}

\begin{figure}[!ht]
\centering
\includegraphics[height=4.8cm]{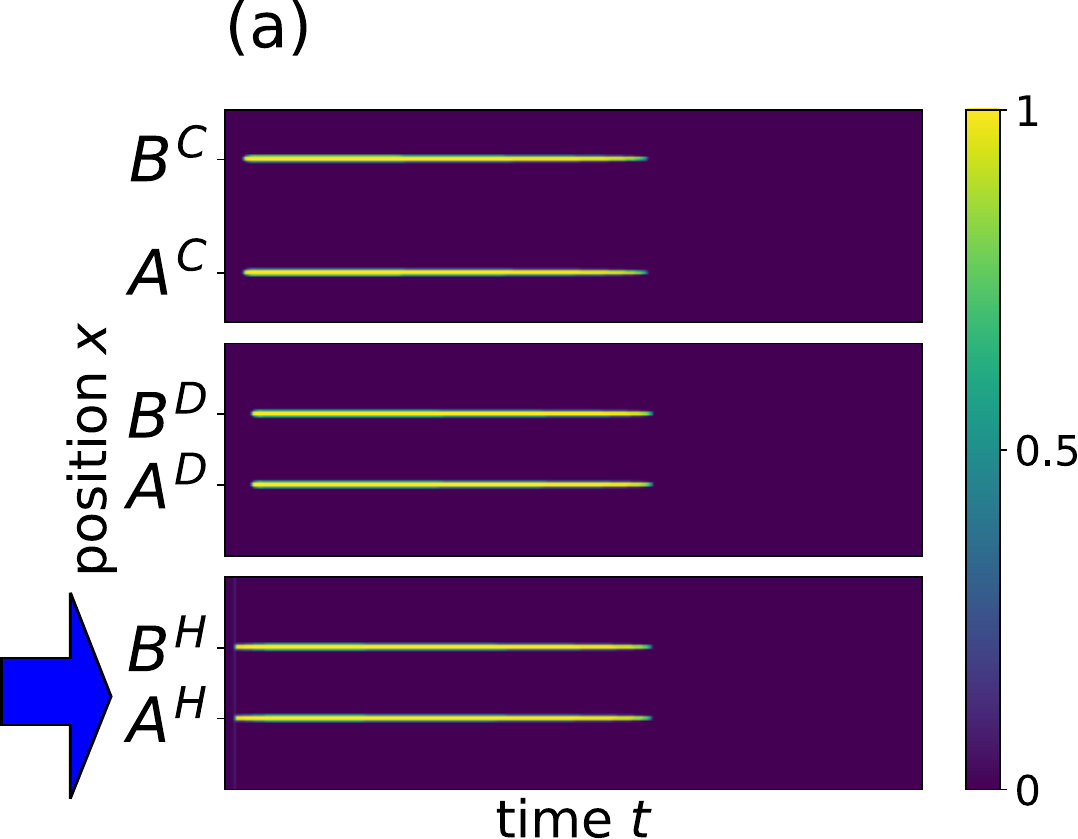}
\includegraphics[height=5cm]{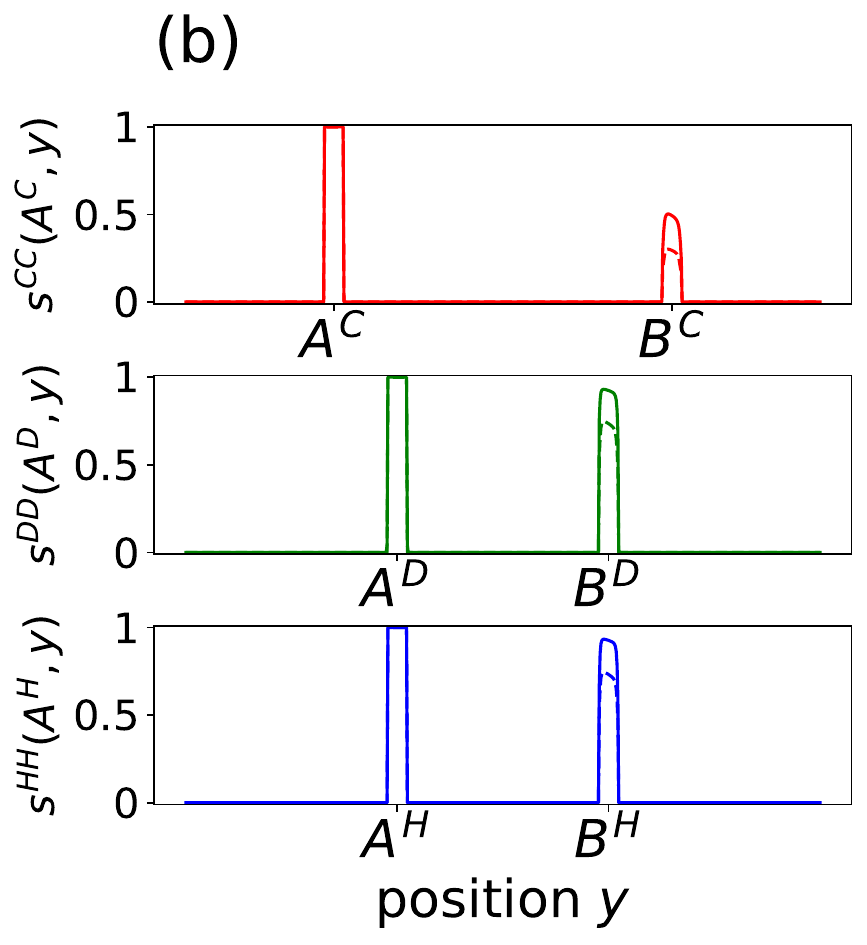}
\includegraphics[height=4cm]{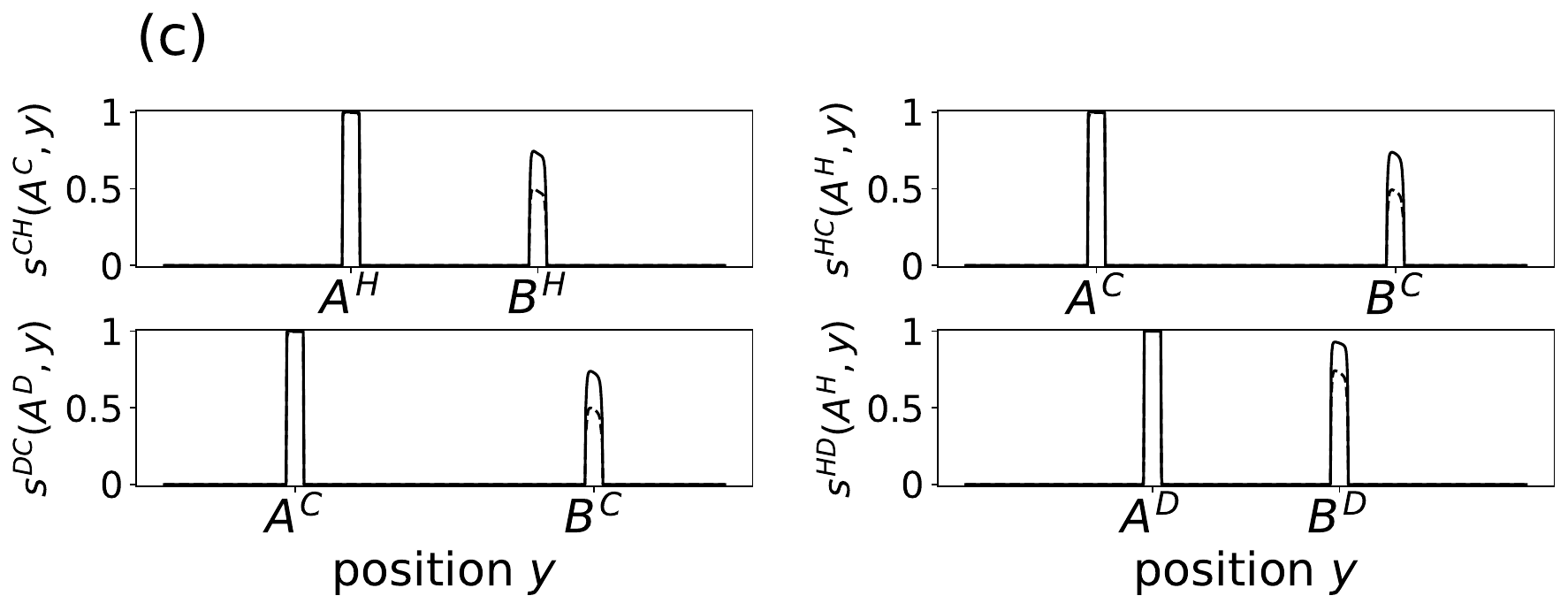}
\caption[The hippocampal replay step allows engrams reactivation]{The hippocampal replay step allows engrams reactivation. \textbf{(a)} Firing rates in the three fields, neocortex (C), dentate gyrus (D) and CA regions (H). We modeled hippocampal reactivation by sending a spatially uniform stimulus on H. All H neurons receive this input, but only those involved in the pattern fire for a long time, because of their non zero learning weights. Activated H neurons lead to firing in C neurons, which then activate D neurons. \textbf{(b)} Intra and \textbf{(c)} inter-field learning weights at the end of the hippocampal replay phase (solid lines) compared to learning weights at the end of the encoding phase (dotted lines). Weights consolidate thanks to pattern reactivation, in particular in the neocortex.}
\label{HR}
\end{figure}

\begin{figure}[!h]
\centering
\includegraphics[height=4.8cm]{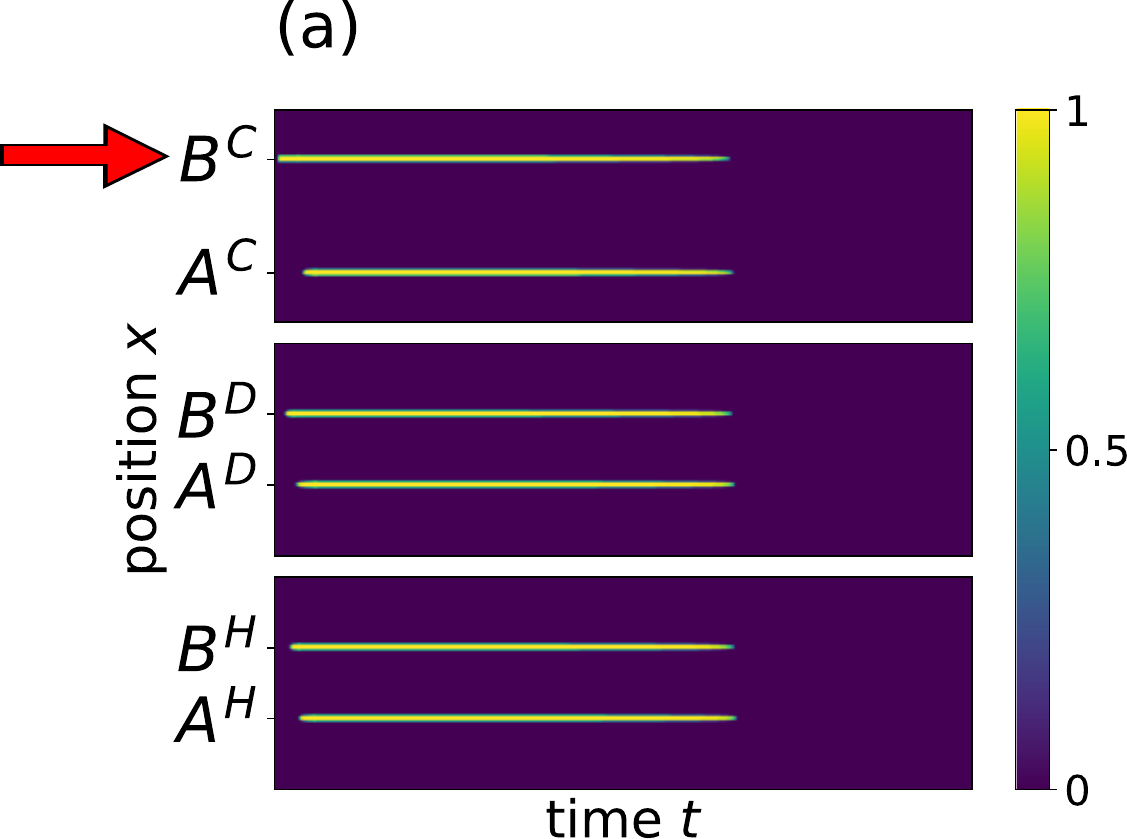}
\includegraphics[height=5cm]{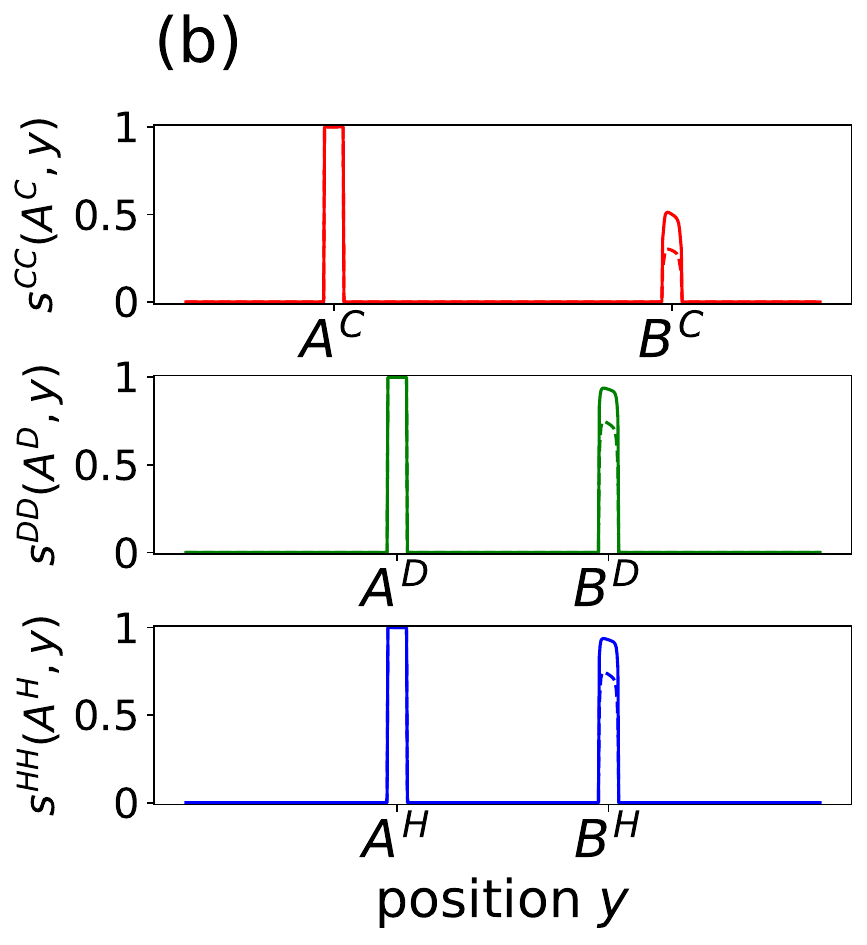}
\includegraphics[height=4cm]{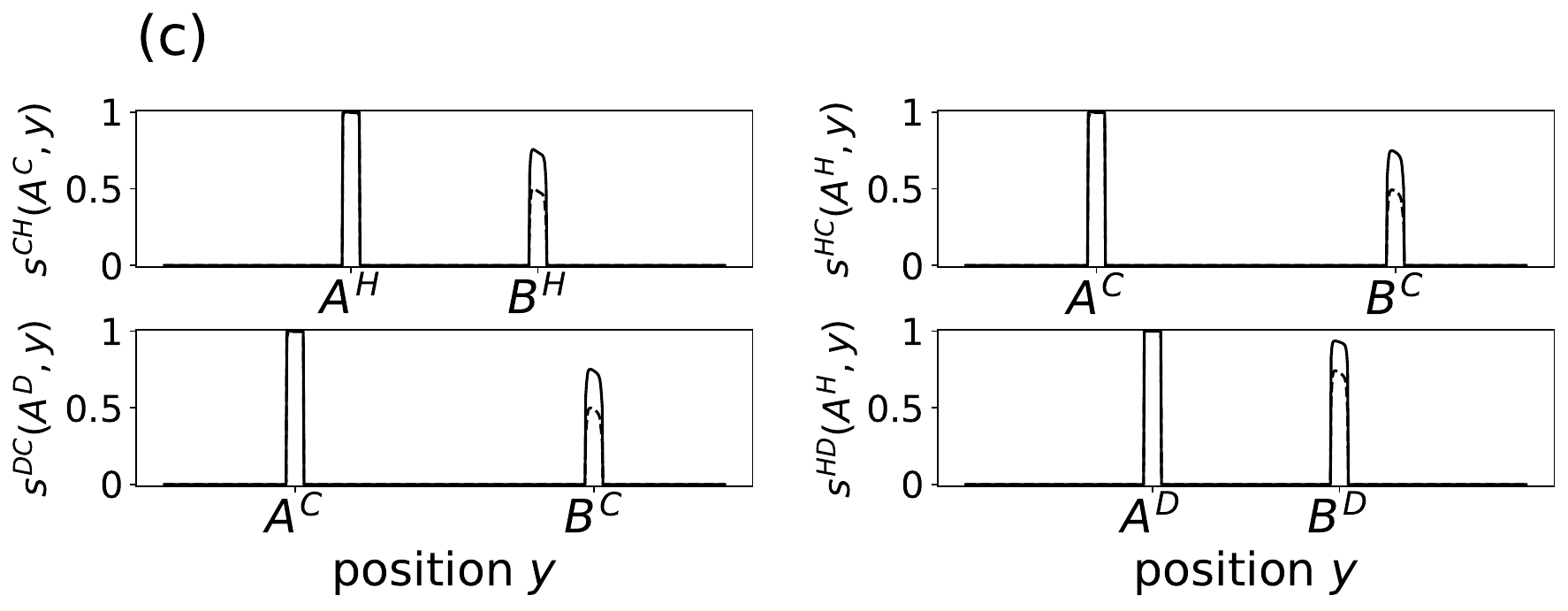}
\caption[The retrieval cue step allows engrams reactivation.]{The retrieval cue step allows engrams reactivation. This step consists in stimulating only the $A^C$ or $B^C$ location of the neocortex (here $B^C$). Retrieval is exhibited by subsequent activation of $A^C$. \textbf{(a)} Firing rates in the three fields, neocortex (C), dentate gyrus (D) and CA regions (H). A signal is sent on only $B^C$ in the neocortical field. The weights between $A^C$ and $B^C$ are still small, C is still dependent on H to recover the whole pattern. Thus, $B^C$ neurons first activate $B^D$ then $B^H$ neurons, which will activate $A^H$ neurons thanks to the strong enough $B^H-A^H$ learning weights. Finally, the $A^H$ neurons activate $A^C$ neurons via the strong $A^C-A^H$ weights, leading to the whole pattern recovery.  \textbf{(b)} Intra and \textbf{(c)} inter-field learning weights at the end of the retrieval cue process (solid lines) compared to learning weights at the end of the encoding phase (dotted lines). Weights are consolidated thanks to pattern reactivation, in particular in the neocortex.}
\label{RC}
\end{figure}

\begin{figure}[!h]
\centering
\includegraphics[height=4.8cm]{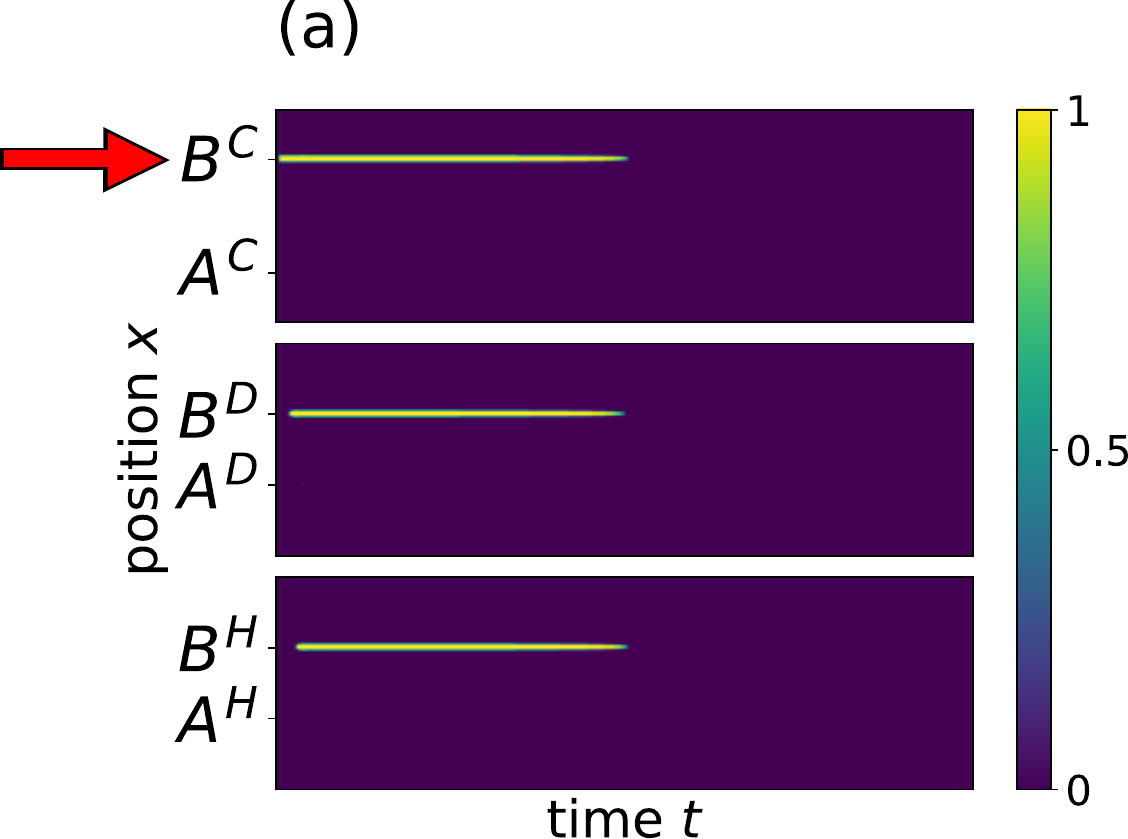}
\includegraphics[height=5cm]{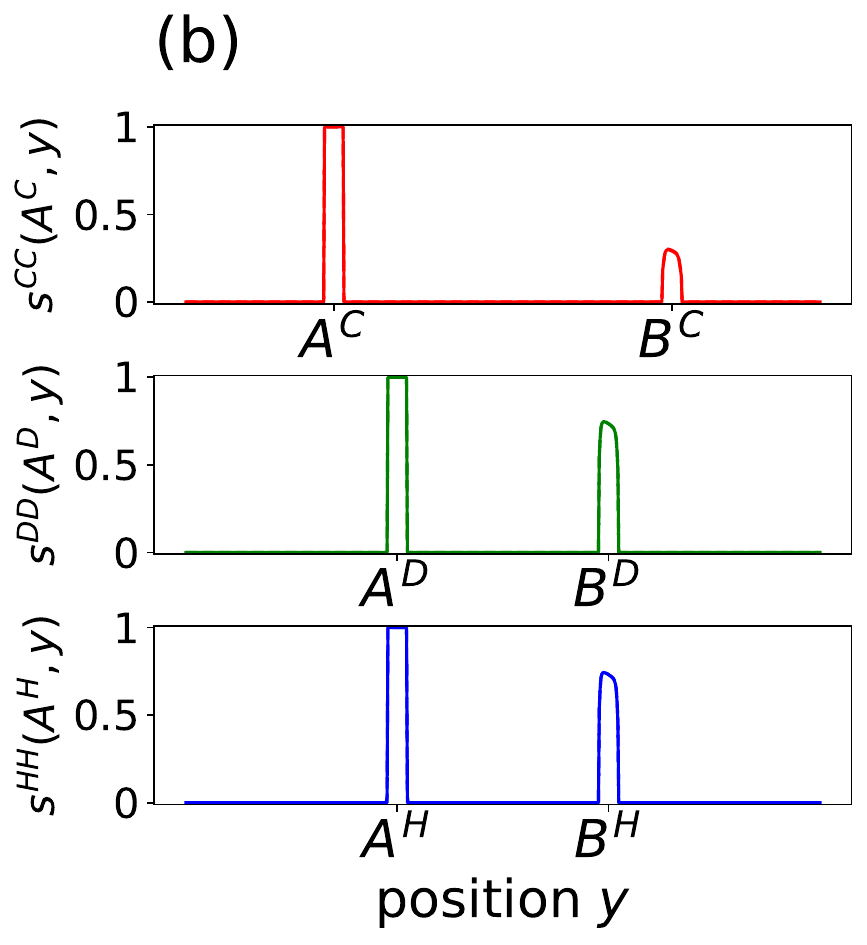}
\includegraphics[height=4cm]{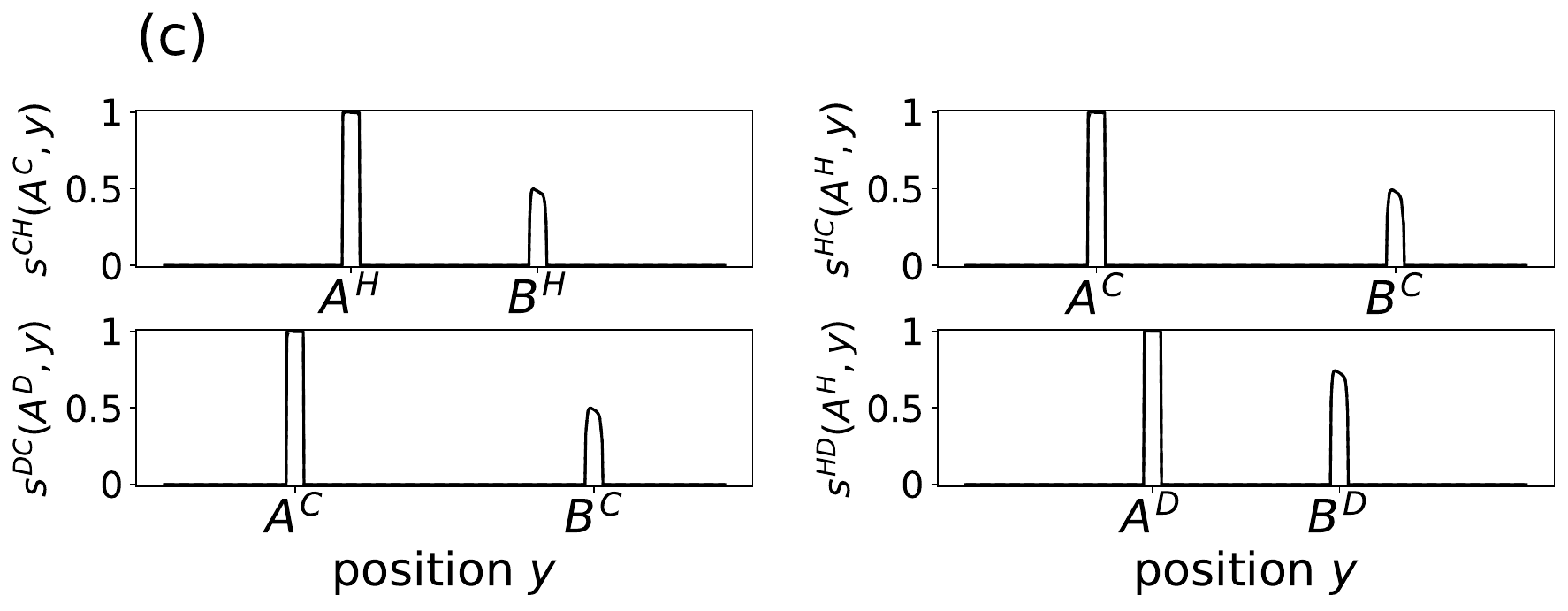}
\caption[Neocortical engram reactivation depends on hippocampal fields during the retrieval cue step. Retrieval cue with hippocampal ``lesion'']{Neocortical engram reactivation depends on hippocampal fields during the retrieval cue steps. The retrieval cue stimulation with hippocampal ``lesion'' is obtained by cancelling the learning weights amplitudes in the hippocampal fields, between them and the neocortical field ($\gamma^{\alpha\beta}=0$ for all $\alpha$,$ \beta$ except $\gamma^{CC}$).  \textbf{(a)} Firing rates in the three fields, neocortex (C), dentate gyrus (D) and CA regions (H). Only the $B^\alpha$ neurons, directly stimulated by the partial retrieval cue, fire. Memory cannot be retrieved without a functional hippocampus. \textbf{(b)} The intra and \textbf{(c)} inter-field learning weights between the $A^\alpha $ and $B^\alpha$ cannot be consolidated in absence of firing of the whole pattern.}
\label{RCtest}
\end{figure}

\begin{figure}[!h]
\centering
\includegraphics[height=4cm]{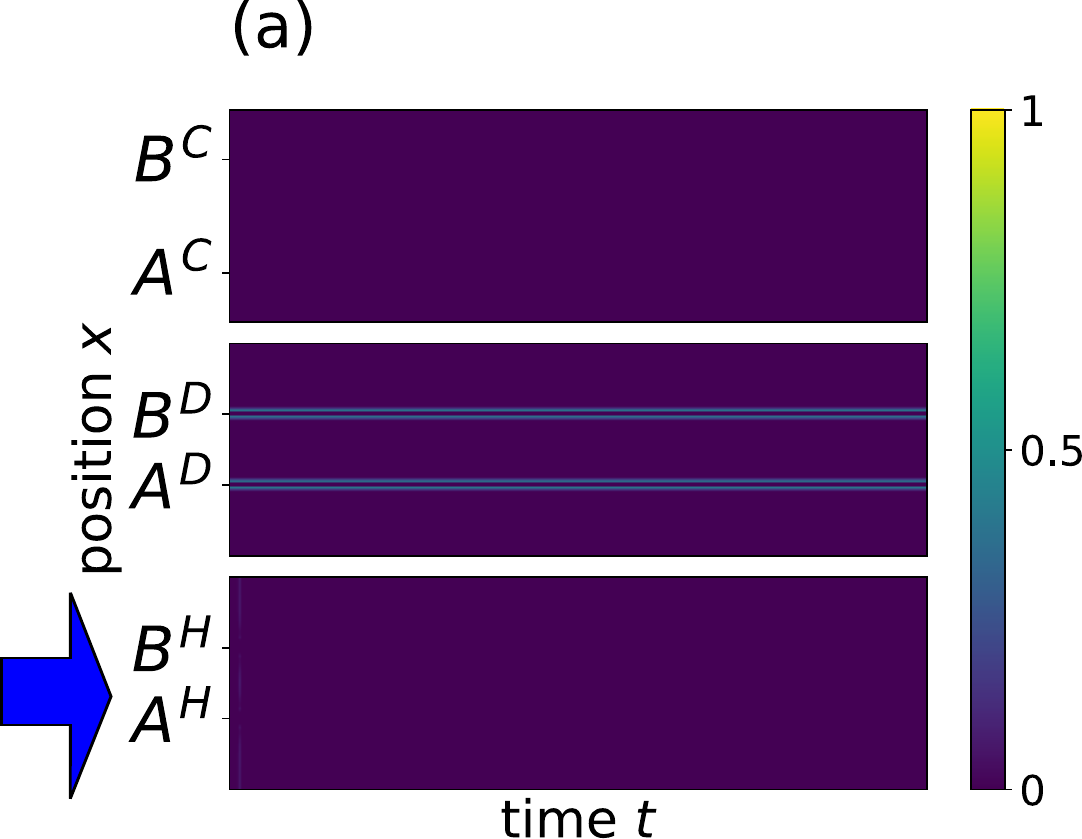}
\includegraphics[height=4cm]{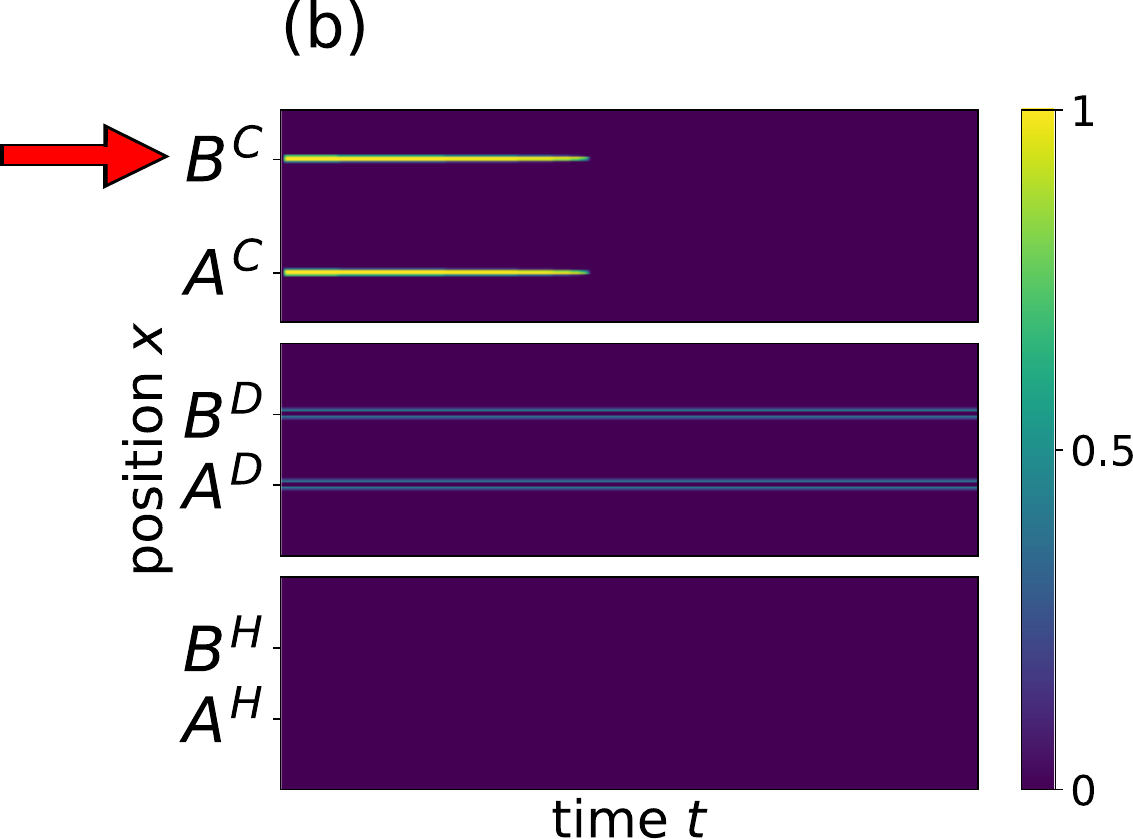}

\includegraphics[height=4cm]{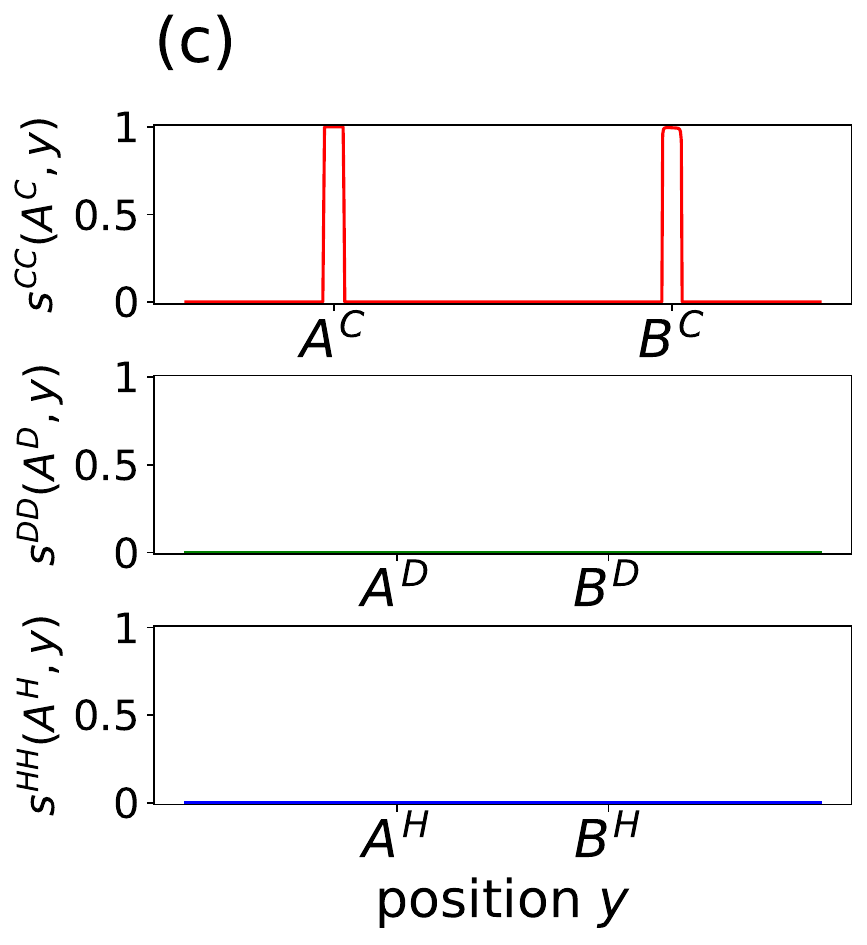}
\includegraphics[height=3cm]{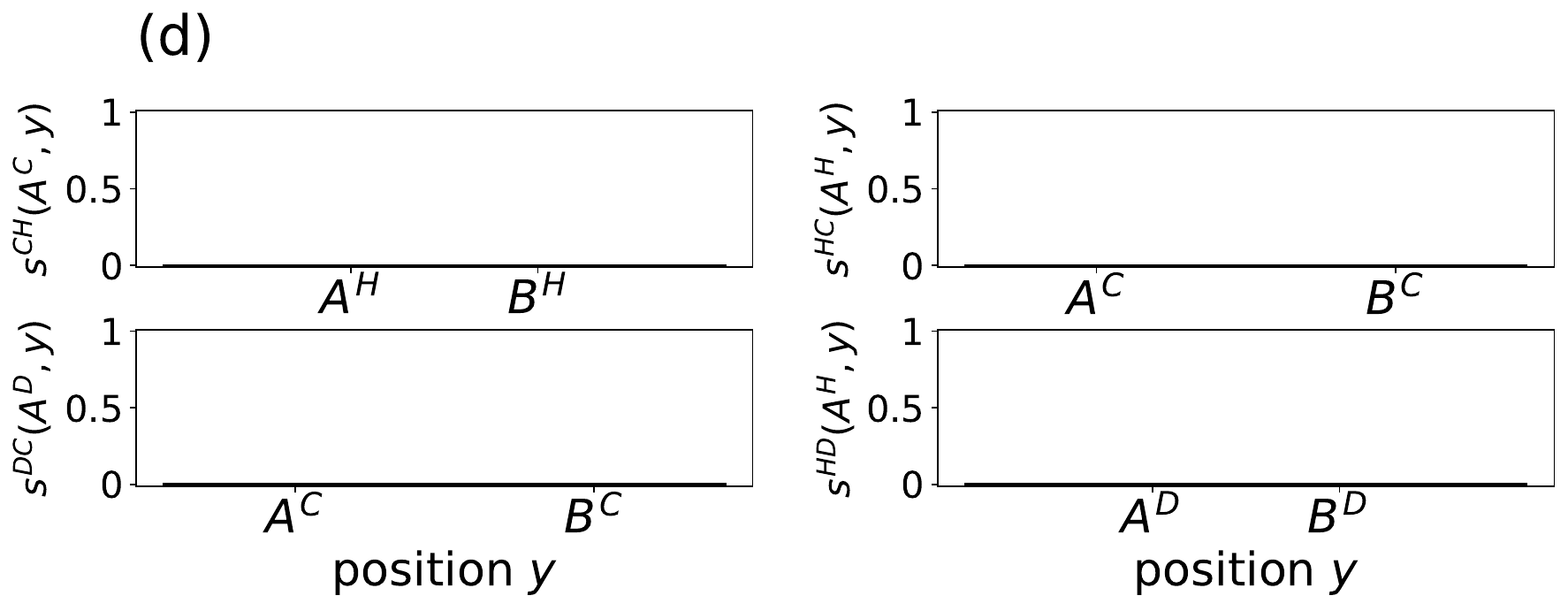}
\caption[Neurogenesis in the dentate gyrus prevents the reactivation of hippocampal engrams]{Neurogenesis in the dentate gyrus prevents the reactivation of hippocampal engrams. \textbf{(a)} Firing rates in the three fields, neocortex (C), dentate gyrus (D) and CA regions (H), during the hippocampal replay phase. Neurons in the neighbourhood of the pattern in the dentate gyrus field are modeled as highly excitable newborn neurons  (lower threshold, see the methods subsection~\ref{methods}). The thin lines in the D field show the continuous firing of those new neurons. As a result, the thresholds of the neurons in the vicinity adapt and stay high. Therefore, the neurons of the pattern stay silent in response to the replay pattern. \textbf{(b)} During neocortical retrieval cues, C neurons fire since there are no direct connections from D to C. However nothing fires in the hippocampal fields. The retrieval cue activates the whole pattern in C independently of the hippocampal fields. \textbf{(c)} Intra and \textbf{(d)} inter-field learning weights after 325 cycles with neurogenesis (solid lines). Learning weights are fully consolidated in the neocortex and have disappeared in hippocampal fields.}
\label{NG}
\end{figure}


\end{document}